\documentclass[journal=jacsat,manuscript=article]{achemso}

\usepackage{chemformula} 
\usepackage[T1]{fontenc} 

\usepackage{amsmath, amssymb}
\usepackage{mathrsfs}
\usepackage{bm}
\usepackage[capitalize]{cleveref}
\usepackage{kotex}
\usepackage{lscape}
\usepackage{xcolor}
\usepackage{lipsum}
\usepackage{subcaption}
\usepackage{multirow}
\usepackage{float} 
\usepackage{graphicx}
\newcommand{\hide}[1]{}

\author{Wanshan Cui}
\altaffiliation{Equal contribution}
\affiliation{Program in Actuarial Science and Financial Engineering, Korea University, Seoul 02841, South Korea}

\author{Yejin Jeong}
\altaffiliation{Equal contribution}
\affiliation{Department of Mathematics, Korea University, Seoul 02841, South Korea}

\author{Inwook Song}
\affiliation{Department of Materials Science and Engineering, Institute of Advanced Materials, Seoul National University, Seoul 08826, South Korea}

\author{Gyuri Kim}
\affiliation{Department of Materials Science and Engineering, Institute of Advanced Materials, Seoul National University, Seoul 08826, South Korea}

\author{Minsang Kwon}
\affiliation{Department of Materials Science and Engineering, Institute of Advanced Materials, Seoul National University, Seoul 08826, South Korea}

\author{Donghun Lee}
\altaffiliation{Corresponding author}
\email{holy@korea.ac.kr}
\affiliation{Department of Mathematics, Korea University, Seoul 02841, South Korea}

\title[An \textsf{achemso} demo]
  {Re-experiment Smart: a Novel Method to Enhance Data-driven Prediction of Mechanical Properties of Epoxy Polymers}

\abbreviations{IR,NMR,UV}

\begin{document}

\hide{
\begin{tocentry}

Some journals require a graphical entry for the Table of Contents.
This should be laid out ``print ready'' so that the sizing of the
text is correct.

Inside the \texttt{tocentry} environment, the font used is Helvetica
8\,pt, as required by \emph{Journal of the American Chemical
Society}.

The surrounding frame is 9\,cm by 3.5\,cm, which is the maximum
permitted for  \emph{Journal of the American Chemical Society}
graphical table of content entries. The box will not resize if the
content is too big: instead it will overflow the edge of the box.

This box and the associated title will always be printed on a
separate page at the end of the document.

\end{tocentry}
}

\begin{tocentry}
\centering
\includegraphics[width=9cm,height=3.5cm,keepaspectratio]{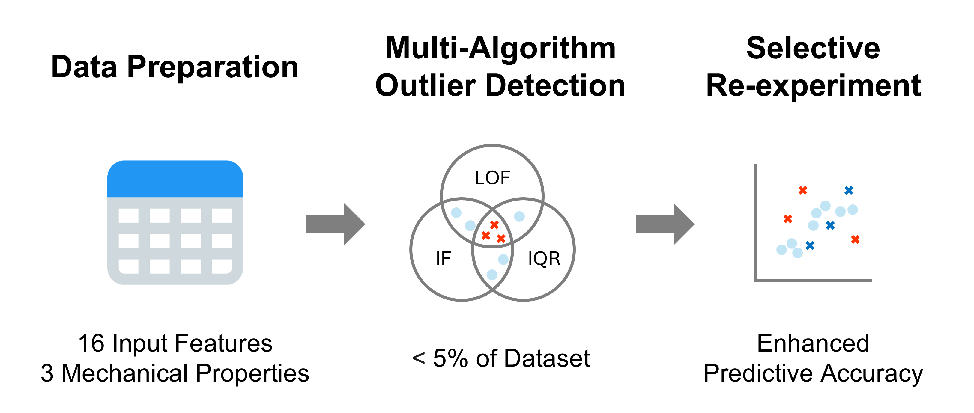}
\end{tocentry}

\begin{abstract}
    Accurate prediction of polymer material properties through data-driven approaches greatly accelerates novel material development by reducing redundant experiments and trial-and-error processes. 
    However, inevitable outliers in empirical measurements can severely skew machine learning results, leading to erroneous prediction models and suboptimal material designs. 
    To address this limitation, we propose a novel approach to enhance dataset quality efficiently by integrating multi-algorithm outlier detection with selective re-experimentation of unreliable outlier cases. 
    To validate the empirical effectiveness of the approach, we systematically construct a new dataset containing 701 measurements of three key mechanical properties: glass transition temperature ($T_g$), tan $\delta$ peak, and crosslinking density ($v_{c}$).
    To demonstrate its general applicability, we report the performance improvements across multiple machine learning models, including Elastic Net, SVR, Random Forest, and TPOT, to predict the three key properties. 
    Our method reliably reduces prediction error (RMSE) and significantly improves accuracy with minimal additional experimental work, requiring only about 5\% of the dataset to be re-measured. 
    These findings highlight the importance of data quality enhancement in achieving reliable machine learning applications in polymer science and present a scalable strategy for improving predictive reliability in materials science.
\end{abstract}

\section{Introduction}
Epoxy adhesives are extensively utilized in a wide range of industries, including automotive, aerospace, and civil engineering, due to their robust adhesion to various substrates, exceptional mechanical properties, and high resistance to heat, corrosion, and chemicals \citep{PROLONGO2006125, JOJIBABU2020102454, AHMADI2019449, SOUSA2018618}. Primarily composed of epoxy resin and hardener (curing agent), epoxy adhesives may incorporate additional additives, such as accelerators and fillers, for modification \citep{article1}. Epoxy adhesives are formulated by subjecting their compositions to a curing process, which can occur at room temperature, elevated temperature, or through alternative methods such as exposure to UV light \citep{JIN20151}.

The properties of epoxy adhesives, attributed to their three-dimensional crosslinked structure, vary depending on the combination of compositions and curing conditions, including cure step, time, and temperature \citep{JIN20151, article1, LAPIQUE2002337}. During the curing process, epoxy resin transforms from liquid to gel and finally to solid states, forming a three-dimensional network structure through crosslinking reactions between epoxide rings and the curing agent, leading to resin polymerization \citep{BARAN202317, article4}. 
Hence, both in research and in production, investigating different epoxy formulations is crucial in producing epoxy adhesive with desired properties.

Dynamic mechanical analysis (DMA) is an essential tool for examining the crosslinked network of an epoxy \citep{LAZZARA2023121} and quantifying its mechanical properties. 
It measures the viscoelastic behavior of polymers by analyzing their response to oscillating stress under varying temperature, time, and frequency conditions \citep{DENARDO2017203}. 
Key measurements include storage modulus (\(E'\), elastic part), loss modulus (\(E''\), viscous part), \(\tan \delta\) (the ratio of \(E''\) to \(E'\)), and glass transition temperature ($T_{g}$, the temperature at which \(\tan \delta\) peak occurs) of tested samples \citep{DENARDO2017203, AKIL2014323}. 
Also, the crosslinking density ($v_{c}$) of a polymer material can be computed with the measured \(E'\) and $T_{g}$ through a specific equation \citep{tianhong2020quantifying, HILL1997235}. 
DMA properties are instrumental in characterizing epoxy adhesives, offering insights into their thermal transitions within polymer-based systems. 
In summary, investigating DMA results provides valuable insights into thermal transitions and final mechanical properties of the polymerization-based adhesive system.

Mechanical properties of an epoxy adhesive depend on both diverse epoxy reagent composition and complex stochastic crosslinking dynamics of epoxy curing process. 
The overwhelming diversity and randomness make it inefficient to perform exhaustive search backed by experiment-based measurements to  achieve the desired epoxy properties \citep{2c00053}.
Hence, complementary approaches for comprehensive understanding and prediction of polymer materials appeared: empirical rules and consulting relevant literature and databases such as Polymer Genome \citep{CHEN2021100595, Kim2018-dh}, and computational methods such as Monte Carlo simulations \citep{Doros} and molecular dynamics (MD) simulation \citep{article6, LI20112920},

Recently, machine learning and deep learning based methods have emerged as another alternative thanks to the advancement of algorithms and computational power \citep{2c00053, CHEN2021100595, STERGIOU2023112031}. 
In the realm of chemistry, various attempts have been made to leverage machine learning for diverse predictions. 
A wide variety of machine learning techniques have been tested for polymer property prediction tasks. 
Machine learning models, such as neural networks and random forests, have recently been successful in predicting various polymer properties, particularly glass transition temperatures measured via DMA \citep{ALCOBACA202092, TANIGUCHI2023100376, doi:10.1021/acs.jcim.1c01031}. 
Notably, ensemble methods like Elastic Net, Random Forest, and Gradient Boosting
\citep{pruksawan2019prediction}, proved effective in predicting epoxy adhesive strength amidst complex data analyses. 
Similarly, Artificial Neural Networks played a pivotal role in accurate predictions regarding material strength and theoretical models \citep{kang2021prediction}.

Despite recent advancements, these data-driven methods share a common weakness -- their performance depends on data quality, as their name suggests. 
In particular, they are highly susceptible to \emph{outliers}, or biased data points that can distort model training and compromise accuracy \citep{article5}.
Researchers inherently strive to mitigate variability in the data collection process to ensure experimental reliability. 
However, it is well known that external factors like human error or experimental variations still arise, potentially affecting the experiment's quality and, consequently, the final polymer's properties \citep{8786096, SMITI2020100306}. 
As repeating error-prone measurements for the entire dataset can be prohibitively expensive, mitigating error effects through extensive repetitions is theoretically feasible but practically unachievable.

Therefore, identifying and handling outliers within datasets made of measurements is vital in maximizing the efficacy of applying data-driven methods to analyze any dataset of expensive empirical measurements, such as DMA-based epoxy polymer properties. 
However, despite the high variability of chemical compounds and polymer synthesis process, previous studies tended to overlook outlier occurrences, as ruling out even more datapoints is discouraged when given a limited dataset of a small sample size. 
Gross error detection methods, suggested by chemistry researchers to spot the occurrence of non-random errors, utilize anomaly detection methods such as Isolation Forest (IF), Local Outlier Factor (LOF), and Interquartile Range (IQR) \citep{dobos2023comparative}.
Developing on this idea, we aim to actively improve the effectiveness of an experiment-based dataset by identifying outlier data points using multiple anomaly detection algorithms \citep{8786096, SMITI2020100306}, and then replacing the outliers by controlled re-experimentation. 

\begin{figure}[th]
  \centering
  \includegraphics[width=1.0\linewidth]{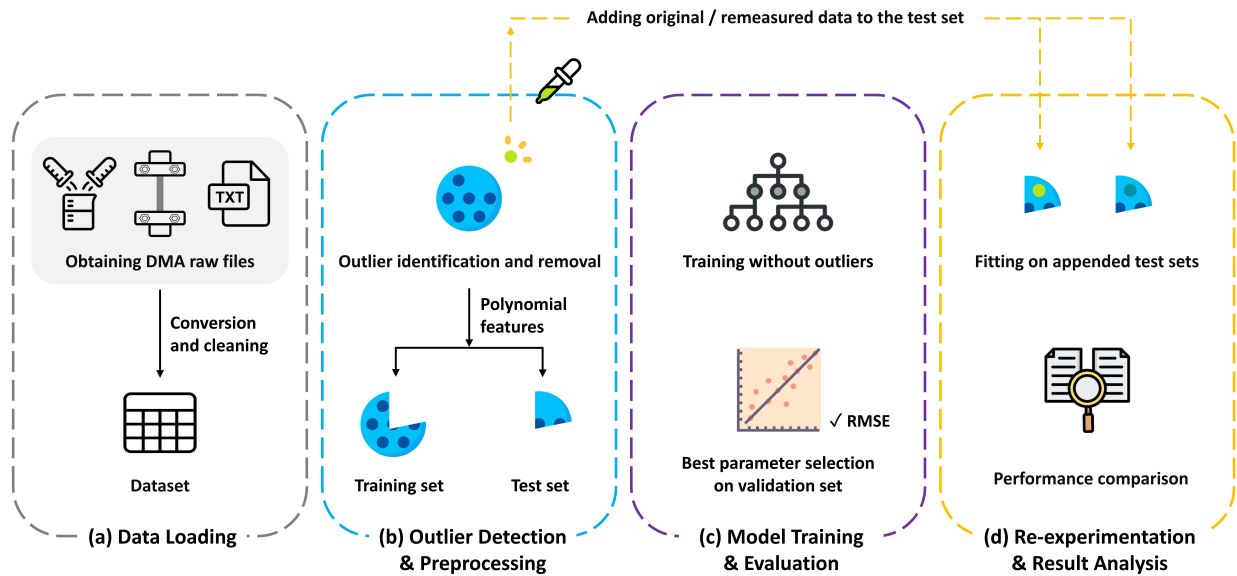}
    \caption{
    Overview of the main contribution, illustrating the four key stages of the proposed method and its workflow. 
    (a) \textbf{Data Loading}: Raw DMA files are collected, converted, and cleaned to form a structured dataset. 
    (b) \textbf{Outlier Detection \& Preprocessing}: Outliers are identified and removed, followed by the generation of polynomial features. The data is then split into training and test sets. 
    (c) \textbf{Model Training \& Evaluation}: Models are trained on data without outliers, with optimal parameters selected based on validation set performance (measured by RMSE). 
    (d) \textbf{Re-experimentation \& Result Analysis}: Original or remeasured data is appended to the test set for further fitting and performance comparison.
    }
    \label{fig:overview}
\end{figure}

In this study, we propose a novel selective re-experimentation approach to mitigate the impact of errors in experimental datasets. 
We use data-driven methods to identify and recondition specific outlier samples, in order to improve the dataset quality for downstream machine learning-based prediction tasks. 
The schematic overview illustrating the proposed re-experimentation approach is shown in \cref{fig:overview}.
This approach enables more accurate prediction of three key epoxy properties: glass transition temperature (Tg), tan $\delta$ peak, and crosslinking density. 
Notably, we demonstrate that significant improvements in predictive performance can be achieved with minimal re-experimentation, involving only a few strategically selected samples. 
Our work uses an exclusively collected dataset comprised of 701 actual experiment-based measurements conducted systematically in a single facility over a set time period to contain the innate variability in the synthesis and the measurement processes. 
Thanks to this unique dataset, the findings from our work have increased credibility as a solid foundation to utilize data-driven methods to predict the properties of thermoset epoxy polymers and to overcome the unavoidable empirical variability of polymer science.

\section{Methods}

\subsection{Theoretical Foundations}

Our goal is to construct a novel method that can improve the data-driven predictive power of the structural properties of thermoset epoxy polymers, by carefully selecting a few re-experiment cases. 
By definition, we assume no knowledge of how the properties depend on various conditions of how epoxy polymers are cured. 
Also, we assume that an arbitrary choice of the machine learning algorithm is made \textit{a priori}, with the goal of predicting the epoxy property $y$ given any experiment setting inputs $\vec{x}$. 

The aforementioned problem can be formulated as a risk minimization problem as follows:
\begin{align}
    \min_{f \in \mathcal{F}} \mathbb{E} \left[ \mathcal{L} \left( f({X}), Y \right) \right] ,  \label{eq:RMgoal}
\end{align}
where the expectation is with respect to the true distributions $\mu_X$ (the distribution of possible experiment inputs ${X}$) and $\mu_{Y|X}$ (the distribution of outcome $Y$ given input ${X}$). 
Also, note that the set of all possible functions learned from the chosen machine learning algorithm is $\mathcal{F}$ and a suitable loss function is $\mathcal{L}$. 
The expectation cannot be computed since the distribution $\mu_{Y|X}$ is unknown, because the knowledge of how epoxy polymer property $Y$ depends on the input variables ${X}$ is unknown by problem definition. 

Therefore, we use the empirical risk minimization (ERM) approach, a purely data-driven, or ``learn-from-data'' approach to solving \cref{eq:RMgoal}. 
Let the measurements from epoxy polymerization experiments comprise a dataset, denoted as $\mathcal{D} := \left\{ \left( \vec{x}_i, \vec{y}_i \right) \right\}$, 
where $\vec{x}_i$ contains information on the monomers and the curing process, and $\vec{y}_i$ contains the properties of the epoxy polymer measured from the $i$-th experiment. 
Given the dataset, we can formulate an ERM problem as follows:
\begin{align}
    \min_{f \in \mathcal{F}} \frac{1}{\left| \mathcal{D} \right|} \sum_{i=1}^{\left| \mathcal{D} \right|} \mathcal{L} \left( f(\vec{x}_i), \vec{y}_i \right) ,  \label{eq:ERMgoal}
\end{align}
whose solution $\tilde{f}: X \rightarrow Y$ is the predictor function learned from the dataset using the given machine learning algorithm that defines $\mathcal{F}$. 
Unlike \cref{eq:RMgoal}, the ERM problem defined as \cref{eq:ERMgoal} can be solved, albeit with a tradeoff: its solution $\tilde{f}$ is likely to be suboptimal. 
This suboptimality is caused by imperfection of the measured dataset $\mathcal{D}$. 

\subsection{Key Idea}

Imperfections in experiment measurements are inevitable, even though all epoxy polymerization experiments and property measurements follow standardized procedures with utmost care. 
The measured properties $y_i$ of the dataset will contain measurement error due to not only the innate variability from polymerization-measurement experimentation itself but also unintentional human error in polymerization or measurement stages. 

\begin{figure}[!h]
  \centering
  \includegraphics[width=1.0\linewidth]{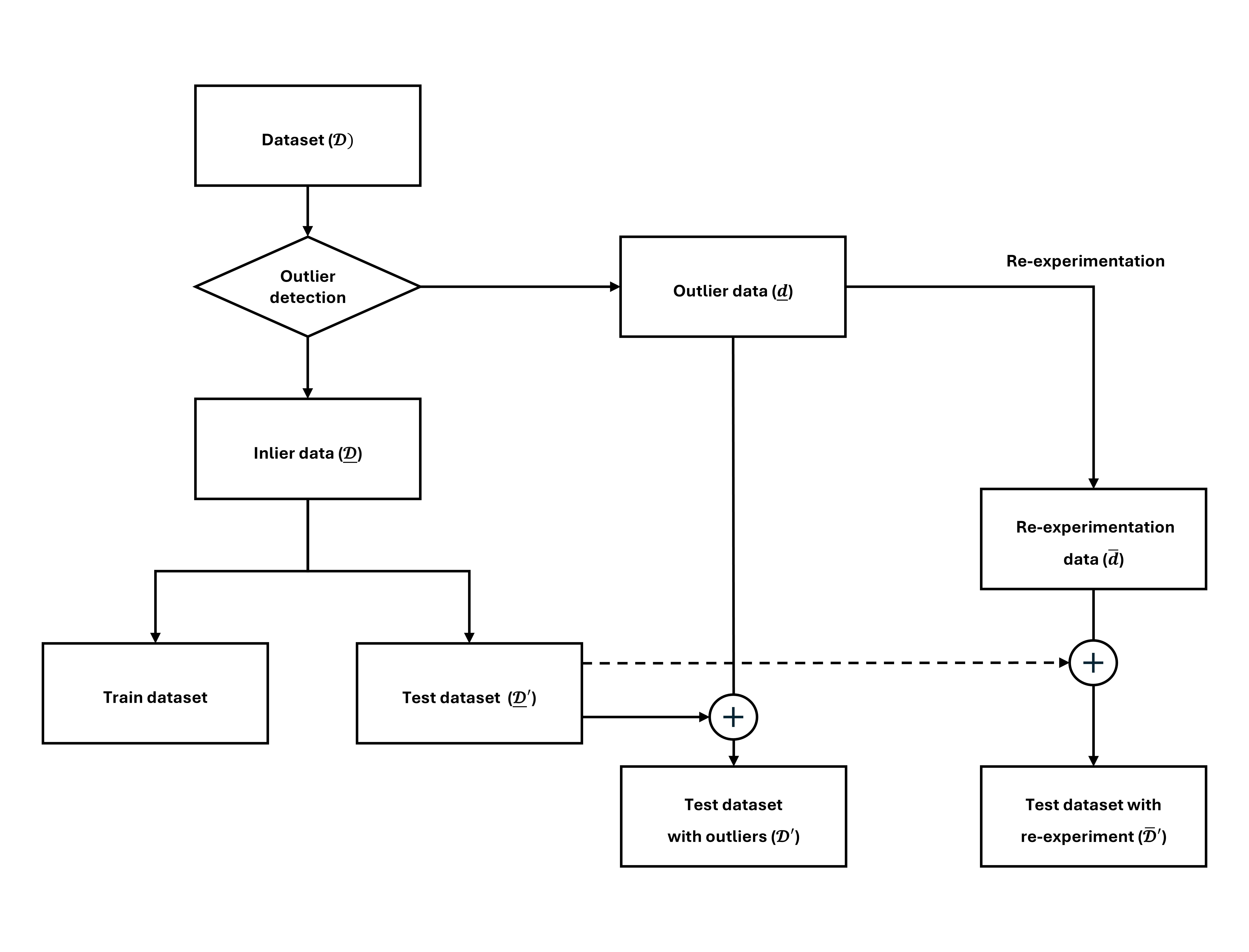}
  \caption{The dataset splitting pipeline begins with outlier detection, which separates the dataset into inliers and outliers. The inlier data is split into training and test datasets, while the outlier data undergoes re-experimentation to generate new data. Finally, the test dataset is combined with the re-experimented outlier data to form the final test dataset with outliers and re-experimentation data.}
  \label{fig:dataset_structure}
\end{figure}

Therefore, we propose a general method to improve machine-learning based prediction of epoxy polymers via repeating only a small fraction of experiments found in $\mathcal{D}$. 
The method comprises two steps: first, determine a subset of outlier data points subject to re-experimentation \emph{before} training a data-driven predictor; second, measure two prediction results -- one with the original data with the outliers and the other with the re-experimented data in which the outliers are replaced with the new measurements -- and quantify the improvement in prediction accuracy. 
To measure the benefit of selective re-experimentation, we design a dataset splitting strategy and corresponding methods to construct the two test sets. 
\cref{fig:dataset_structure} shows a flowchart outlining the splitting strategy. 

The splitting and training steps are as follows: 
First, we construct an outlier detection method $g$ to select the subset $\underline{d}$ from the original dataset $\mathcal{D}$. 
After removing those data points, we get a reduced dataset $\mathcal{\underline{D}}:=\mathcal{D} \setminus \underline{d}$. 
Then, we split the reduced dataset into the common test set $\mathcal{\underline{D}}'$ and the training set, in which no outliers are included. 
We only use the training set to construct the predictor, which is the optimal solution for the ERM problem formulated in \cref{eq:ERMgoal}. 

To test the predictor, we construct two test sets. 
The first test set, which corresponds to the usual case of containing the outliers $\underline{d}$, is defined as $\mathcal{D}' := \mathcal{\underline{D}}' \cup \underline{d}$. 
We perform selective re-experimentation and denote the set of newly measured data points as $\overline{d}$, which can be used to replace the outliers $\underline{d}$ one-by-one. 
The second test set, which corresponds to the outlier-replaced-with-re-experimentation case, is defined as $\mathcal{\overline{D}}' := \mathcal{\underline{D}}' \cup \overline{d} = \left( \mathcal{D}' \setminus \underline{d} \right) \cup \overline{d}$. 

We hypothesize that such selective re-experimentation can improve the quality of the dataset with only a tiny fraction of the cost compared to replicating the entire set of measurements. 
We expect to demonstrate the improvement by showing a range of machine-learning algorithms $f$ trained on the common training set will have significantly better prediction accuracy with the partially re-experimented test set $\mathcal{\overline{D}}'$ than with the original test set $\mathcal{{D}}'$. 

\hide{
\begin{figure}[th]
  \centering
  \includegraphics[width=1.0\linewidth]{overview_v2.png}
  \caption{Pipeline overview}
\end{figure}
}

\begin{figure}[th]
  \centering
  \includegraphics[width=1.0\linewidth]{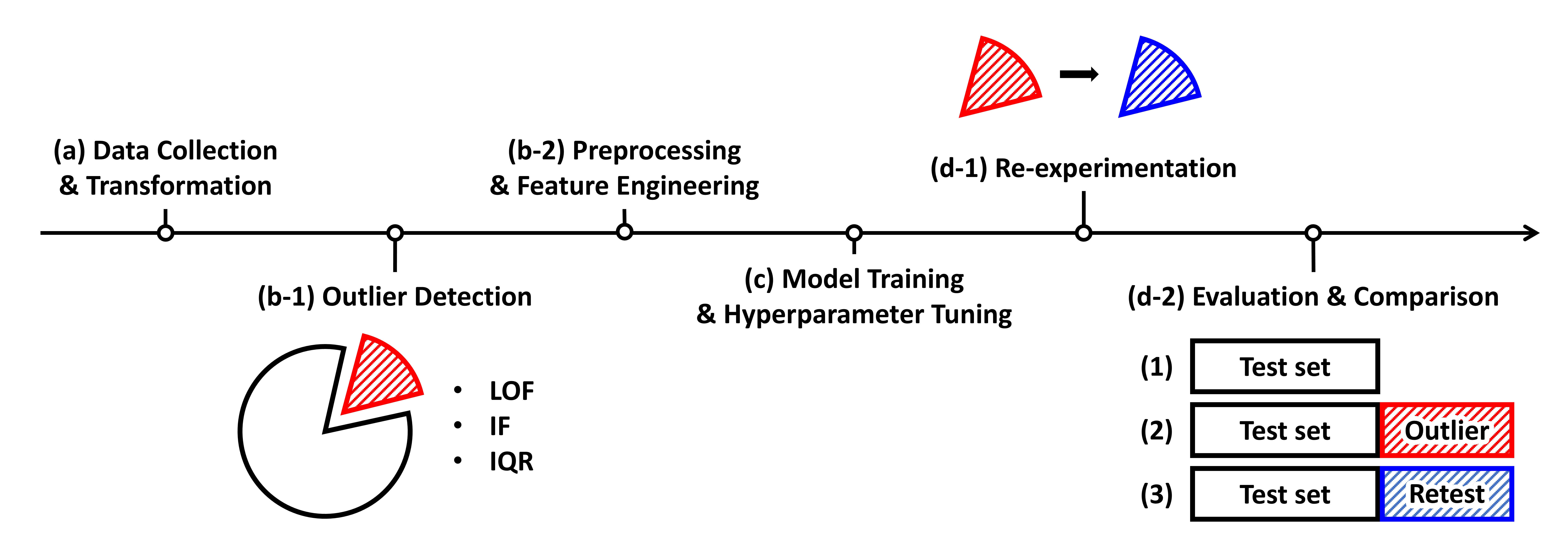}
  \caption{Overview of the machine learning pipeline for epoxy property prediction. The workflow consists of six key stages: (a) Data Collection \& Transformation, (b-1) Outlier Detection using LOF, IF, and IQR algorithms, (b-2) Preprocessing \& Feature Engineering, (c) Model Training \& Hyperparameter Tuning, (d-1) Re-experimentation to replace outliers with refined samples, and (d-2) Evaluation \& Comparison using three test scenarios: (1) Test set, (2) Test set with outliers (red), and (3) Test set with re-experimented data (blue).}
  \label{fig3}
\end{figure}

\subsection{Experiment Construction and Validation}
In this study, multi-input single-output prediction models are developed for each of the three properties of thermoset epoxy to be measured and predicted. 
The process for building machine learning prediction models of each of the target properties follows the streamlined pipeline outlined in \cref{fig3}.

Firstly, we import and transform raw data from the Dynamic Mechanical Analyzer into structured data suitable for analysis. 
Next, we employ outlier detection techniques to filter outliers based on the values of each property and preprocess the remaining dataset. 
During the preprocessing, we perform feature engineering by constructing a feature matrix containing polynomial combinations of the input variables up to a specified degree. 
The degree of the polynomial is a hyperparameter, which can be optimized during model training to enhance performance while accommodating the distinct characteristics of the regression models. 
We vary the hyperparameter from one to three. 
In the third step, as part of model training and parameter tuning, we implement 5-fold cross-validation to assess the robustness and generalization ability of the selected models. 
Additionally, we focus on minimizing the root mean square error (RMSE) metric on the validation set to optimize performance. 
Finally, we re-execute the experiment under identical conditions to obtain new property observations for the previously identified outliers. 
The resulting re-experimentation data ($\overline{d}$) is then integrated into the existing test set ($\mathcal{\underline{D}}'$) used in the third step. 
Subsequently, we evaluate this appended test set ($\mathcal{\overline{D}}'$) using the best-performing models identified earlier to assess any alterations in predictive outcomes compared to the original test scores. 

To validate the efficacy of the outlier detection process, we re-integrate the original outlier data ($\underline{d}$), collected before re-experimentation, into the test set ($\mathcal{\underline{D}}'$), thus creating another appended test set ($\mathcal{D}'$). 
We subsequently conduct a comparative analysis employing the same models to evaluate scores derived from the three test sets introduced, utilizing root mean square error (RMSE) and coefficient of determination ($R^{2}$) as evaluation metrics.

By following these sequential steps, we ensure that we adopt a robust and systematic approach toward data preprocessing, model training, and evaluation, thereby facilitating comprehensive analysis and interpretation of experimental results.

\subsection{Outlier Detection Algorithms}
Due to the inherent variability of output values even under identical experimental conditions, manually detecting errors during experiments is challenging. 
Thus we employ three different methods: Local Outlier Factor (LOF), Isolation Forest (IF), and Interquartile Range (IQR).
LOF \citep{10.1145/335191.335388} identifies outliers based on the relative density of data points compared to their neighbors, with lower density points flagged as potential outliers.
IF \citep{4781136} is an algorithm that recursively partitions the data to find outliers. It randomly selects features to partition the data, and data points with fewer partitions are considered outliers.
IQR method \citep{siegel1996statistics} identifies outliers using the interquartile range (IQR), defined as the difference between the first quartile ($Q1$) and the third quartile ($Q3$), representing the middle 50\% of the data. Data points are considered outliers if they fall below a lower bound ($Q1 - \textit{multiplier} \times \text{IQR}$) or above an upper bound ($Q3 + \textit{multiplier} \times \text{IQR}$). The multiplier (commonly set to 1.5) determines the sensitivity of the method to detecting outliers.

With the aim of improving the accuracy of property predictions, the outliers that significantly deviate from the mainstream data distribution is removed. 
For each property to be predicted, outlier detection was conducted using the three aforementioned methods. 
In particular, the target property values were applied to each outlier detection algorithm, and the intersection of the data identified by each method was taken to select the final outlier dataset.

\subsection{Regression Algorithms for Validation of Effectiveness}
\label{sec:Regression Algorithms}

To assess the empirical effectiveness of our approach, we selected eight supervised machine learning algorithms, categorized into four main types. 
The algorithms and their relevant characteristics are briefly summarized below.

\textit{Linear models: Elastic Net and Bayesian Ridge} \citep{Zou} are computationally efficient and intuitive, offering good interpretability, which helps to understand the impact of independent variables on dependent variables. They are robust to noise and, in certain cases, can outperform more complex nonlinear models when dealing with noisy data or outliers. As a result, they are valuable tools in statistical analysis and predictive modeling. 

\textit{Nonlinear models: Support Vector Regression (SVR) and K-Nearest Neighbors (KNN)} \citep{article7} excel at capturing nonlinear relationships due to their flexibility in modeling complex data patterns. SVR maps the input data into a higher-dimensional space, enabling it to learn more complex decision boundaries. KNN relies on the similarity between the data points to make predictions, effectively capturing local patterns. When the underlying data relationships are nonlinear, these models often outperform linear models in predictive accuracy, leading to more reliable predictions.

\textit{Ensemble Models: Random Forest, Gradient Boosting, and LightGBM} \citep{8320256, Bentejac2021-wl} are ensemble methods that generate decision trees and learn by branching out from each tree. Random Forest aggregates predictions from randomly generated trees, offering robustness to overfitting and enhancing predictive accuracy through aggregation. Gradient Boosting builds trees sequentially, focusing on the residuals from previous trees, thereby refining predictions through successive corrections. LightGBM employs leaf-wise splitting to prioritize more balanced node splits, enhancing both speed and efficiency in large-scale datasets.

\textit{AutoML: Tree-based Pipeline Optimization Tool (TPOT)} \citep{pmlr-v64-olson-tpot-2016} streamlines the model optimization process by automatically exploring various pipeline configurations and selecting the best-performing models. This comprehensive approach increases efficiency in terms of time and prediction compared to traditional machine learning methods.

\subsection{Dataset Preparation}

\begin{figure}[th]
  \centering
  \includegraphics[width=1.0\linewidth]{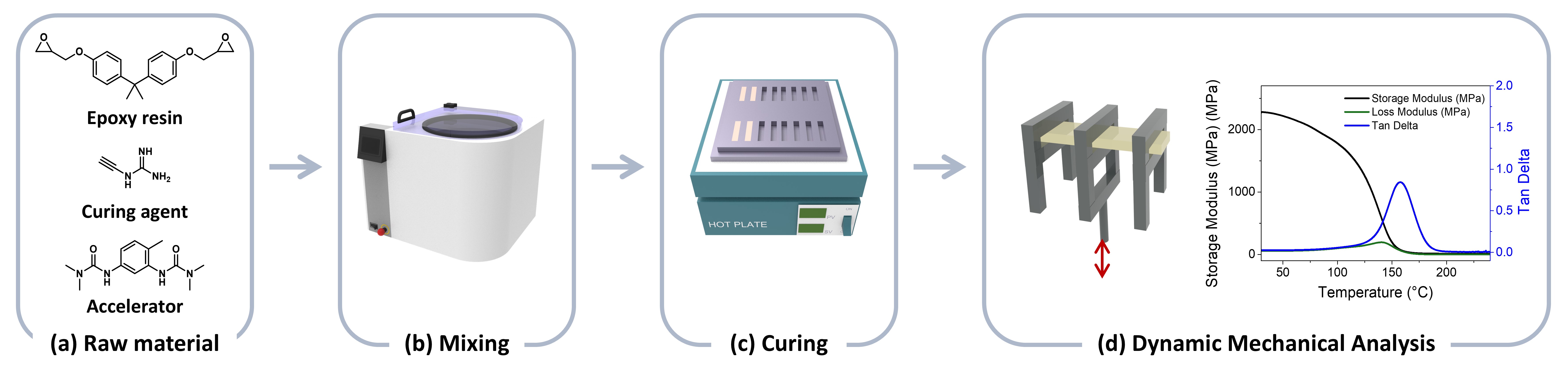}
  \caption{
  Schematic representation of the epoxy sample preparation and measurement process:
  (a) \textbf{Raw Materials}: Epoxy resin(YD-128), curing agent, and accelerator.
  (b) \textbf{Mixing}: The raw materials are thoroughly mixed to ensure homogeneity.
  (c) \textbf{Curing}: The mixture is poured into a mold placed on a hot plate and thermally cured under controlled conditions.
  (d) \textbf{Dynamic Mechanical Analysis (DMA)}: Data analysis is performed using DMA measurements.}
  
  \label{fig4}
\end{figure}

As shown in \cref{fig4}, the overall process begins with material preparation and mixing, followed by molding, curing, and analysis. Using a vacuum paste mixer (THINKY Corporation, ARV-310P), measured amounts of epoxy resin, curing agent, and accelerator were physically and uniformly mixed. The mixing process followed a standardized procedure: initially, the components were mixed at 1500 rpm and 50 kPa for 150 seconds, followed by a second mixing phase at 1300 rpm and 30 kPa for 90 seconds.
After mixing, the prepared mixture was poured into molds designed for DMA specimens. The molds were placed on a hot plate set to the curing temperature. The dimensions of the samples were 12.3 mm x 3.2 mm x 60 mm.
The cured samples were then analyzed using a DMA (TA instrument, Q800). The analysis was conducted at a frequency of 1 Hz and a strain of 0.1\%. The temperature was increased from 30 °C to 220 °C at a rate of 5°C/min to measure the properties.

The dataset comprises 701 data points from experiments, with 16 independent variables selected as features to predict the properties measured from the experiments. 
These variables are derived from the chemical compounds and the curing process variables used in the production of epoxy specimens. 
Specifically, the compositions include epoxy resin, curing agent, and accelerator, which are mixed in various proportions, accounting for ten independent variables. 
For example, the data includes the weight (g) and weight percentage (\%) of each ingredient material. 
Bisphenol A diglycidyl ether (DGEBA, YD-128, KUKDO Chemical) is used as the epoxy resin in all experiments, while carboxyl-terminated butadiene acrylonitrile (CTBN) modified epoxy resin (KR-450, KUKDO Chemical) and core-shell rubber (CSR) modified epoxy resin (KDAD-7101, KUKDO Chemical) are mixed in 30\% and 29\% of the total experimental data, respectively. 
Specifically, these resins are mixed in at most two combinations.  
The curing agent, Dicyandiamide (DYHARD\textsuperscript{®} 100S, AlzChem), and the accelerator, 1,1'-(4-methyl-m-phenylene)bis(3,3'-dimethylurea) (DYHARD\textsuperscript{®} UR500, AlzChem), are employed. 
As all compositions undergo mixing under identical conditions, factors such as rotational speed, pressure, and mixing time during the compounding process are not considered. 
Subsequently, the compositions undergo one to three steps of curing processes, with variations in curing temperature (ranging from 90°C to 160°C) and time (ranging from 0.5 hours to 3 hours per step), totaling six variables.

The dependent variables to be predicted are the three DMA property values: glass transition temperature $T_g$, tan $\delta$ peak, and crosslinking density $v_{c}$, observed from the same experiment. These dependent variables share identical independent variables.  
For consistency, we treat some variables as 0 when certain resins are not involved in the experiment or when fewer than three curing steps occur, as they are not applicable in those cases.

\section{Results}

\subsection{Empirical Validation of Re-experimentation Based Dataset Quality Enhancement}
\begin{figure}[H]
    \centering
    \begin{subfigure}[b]{0.32\textwidth}
        \centering
        \includegraphics[width=\textwidth]{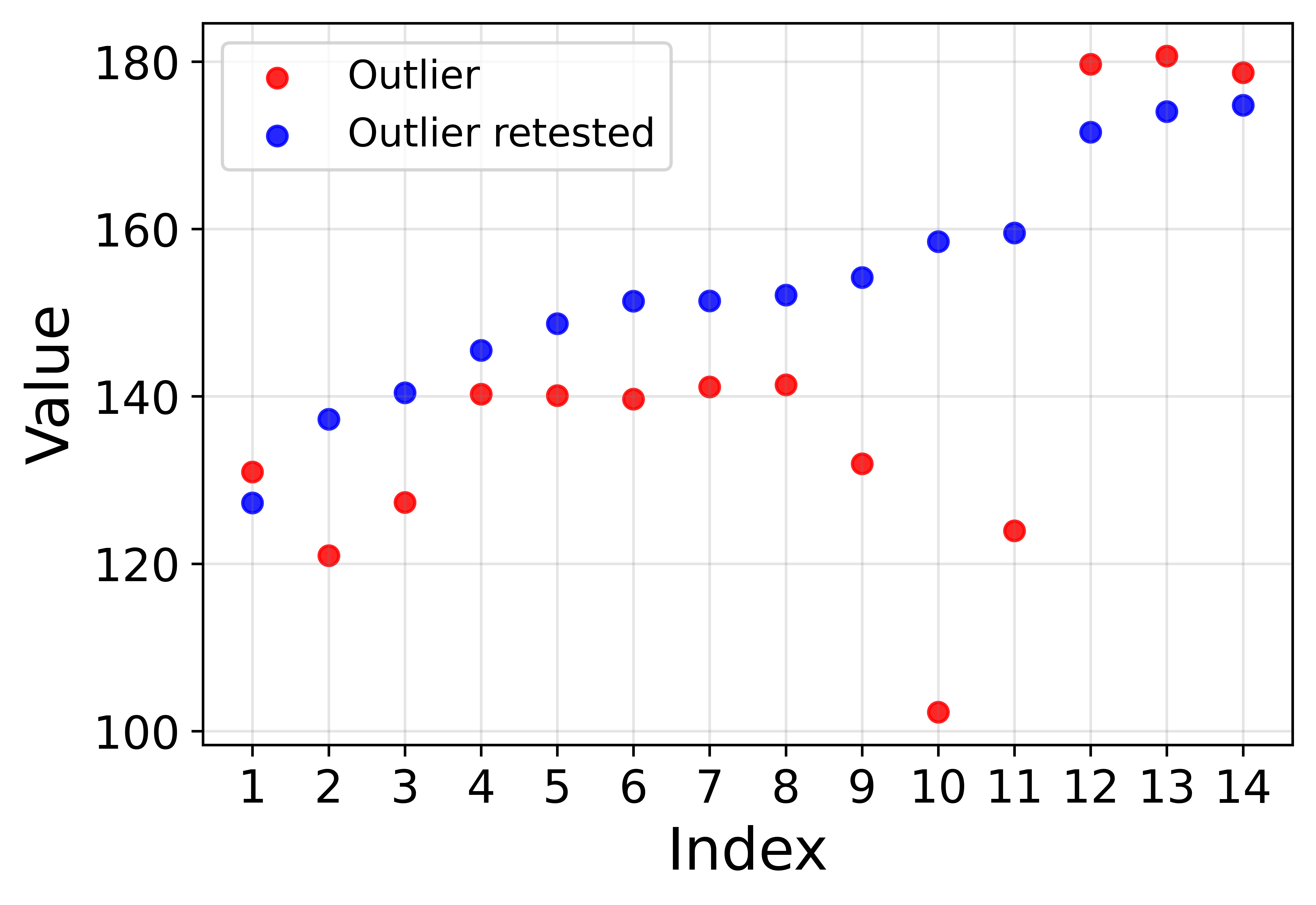}
        \caption{Glass transition temperature ($T_{g}$)}
    \end{subfigure}
    \hfill
    \begin{subfigure}[b]{0.32\textwidth}
        \centering
        \includegraphics[width=\textwidth]{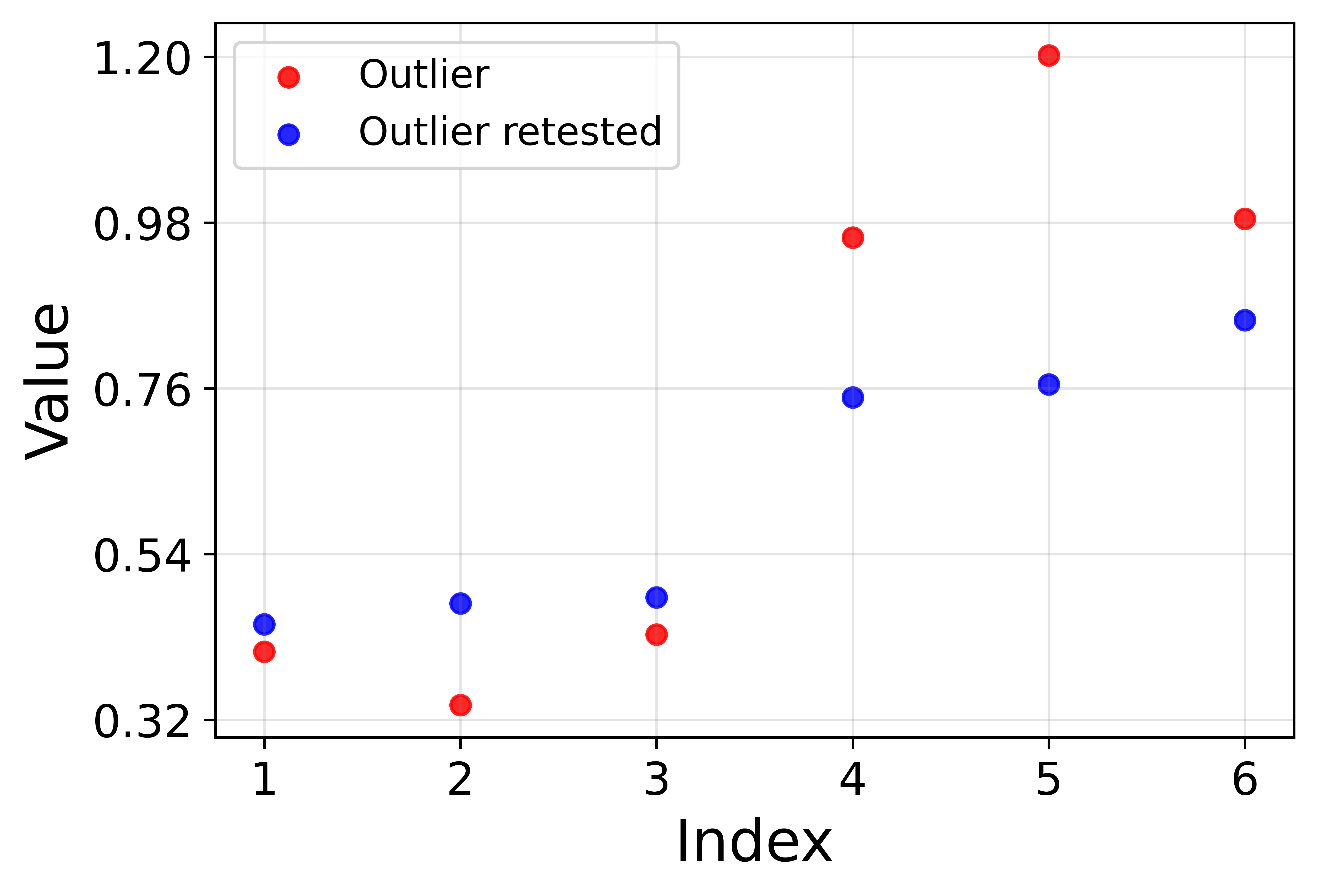}
        \caption{Tan $\delta$ peak}
    \end{subfigure}
    \hfill
    \begin{subfigure}[b]{0.333\textwidth}
        \centering
        \includegraphics[width=\textwidth]{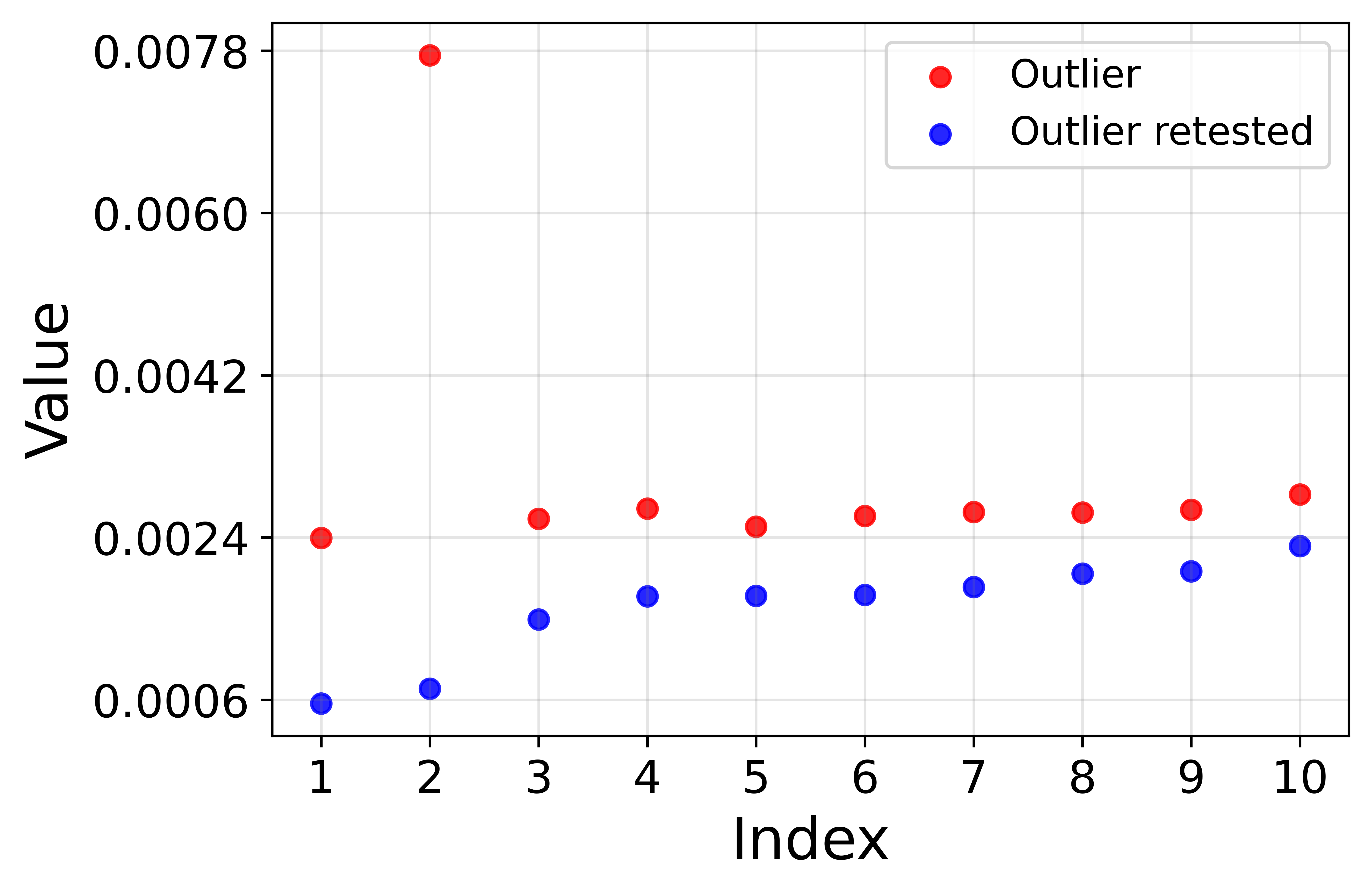}
        \caption{Crosslinking density ($v_{c}$)}
    \end{subfigure}
    \caption{Comparison of initial measurements identified as outliers and measurements after re-experimentation for each property. The index number indicates the order of property values after re-experimentation, arranged from smallest to largest.}
    \label{fig:re-experimentation}
\end{figure}

Following outlier detection, we identified 14, 6, and 10 outliers in the glass transition temperature ($T_g$), tan $\delta$ peak, and crosslinking density ($v_{c}$), respectively. 
We conducted re-experiments under identical conditions (composition ratio, curing temperature, and time) to remeasure these property values.  
As shown in \cref{fig:re-experimentation}, where the actual value of each outlier and its re-experimented value are plotted at the same index, the re-experimented results deviate substantially from the initially detected outliers, indicating that the outliers may have arisen from experimental uncertainties or measurement errors.  

Eight machine learning algorithms, detailed in \cref{sec:Regression Algorithms}, were employed for regression analysis on the epoxy polymer dataset.  
To ensure robustness, we repeated model training with ten random seeds to mitigate variations in train-test data splits. 
The averaged RMSE and $R^2$ values, serving as evaluation metrics, are summarized in \cref{tab:total-table}.

No single model consistently outperformed the others across all configurations.  
The predictive performance of each model varied depending on the property being predicted and the dataset type (inlier, outlier, or re-experimented). 
This indicates that model selection and performance are highly context-dependent, requiring careful evaluation for each scenario. 

Among the three properties, the prediction performance for the tan $\delta$ peak is relatively superior, with consistently lower RMSE values and higher $R^2$ scores across most models. 
This suggests that the tan $\delta$ peak exhibits more predictable patterns or stronger correlations with input features than the other properties. 
When comparing the performance for the two appended test sets, notable differences are observed in the predictive performance of glass transition temperature ($T_g$) and crosslinking density ($v_c$) across different models. 
These differences highlight the potential variability introduced during experiments and underscore the importance of robust model validation to account for experimental inconsistencies or variations in data quality.

\begin{landscape}
\centering
\begin{table}
  \small
  \caption{Mean prediction scores (arithmetic averages) for different models across ten data split seeds. The appended test set, denoted as $\mathcal{\overline{D}}'$ or $\mathcal{D}'$, refers to a renewed test set obtained by adding either the re-experimentation data or the outliers to the test set.}
  \label{tab:total-table}

\begin{center}
  \begin{subtable}{\textwidth}
    \centering
    \small
    \resizebox{0.95\textwidth}{!}{
    \begin{tabular}{lcccccccccccccc}
\hline
\multicolumn{1}{c}{} & \multicolumn{7}{c}{RMSE} & \multicolumn{7}{c}{$R^{2}$} \\ \cline{2-15} 
\multicolumn{1}{c}{} & \multicolumn{1}{c}{} & \multicolumn{1}{c}{} & \multicolumn{5}{c}{Appended test set} & \multicolumn{1}{c}{} & \multicolumn{1}{c}{} & \multicolumn{5}{c}{Appended test set} \\ \cline{4-8} \cline{11-15} 
\multicolumn{1}{c}{\multirow{-3}{*}{Model}} & \multicolumn{1}{c}{\multirow{-2}{*}{Training set}} & \multicolumn{1}{c}{\multirow{-2}{*}{Test set}} & \multicolumn{2}{c}{w/ re-expt. ($\mathcal{\overline{D}}'$)} & \multicolumn{2}{c}{w/ outliers ($\mathcal{D}'$)} & \multicolumn{1}{l}{$\mathcal{\overline{D}}$ vs. $\mathcal{D}'$} & \multicolumn{1}{c}{\multirow{-2}{*}{Training set}} & \multicolumn{1}{c}{\multirow{-2}{*}{Test set}} & \multicolumn{2}{c}{w/ re-expt. ($\mathcal{\overline{D}}'$)} & \multicolumn{2}{c}{w/ outliers ($\mathcal{D}'$)} & \multicolumn{1}{l}{$\mathcal{\overline{D}}'$ vs. $\mathcal{D}'$} \\ \hline
Elastic Net & 3.28142 & {\color[HTML]{333333} 3.79943} & 4.68546 & {\color[HTML]{CC0000} (+0.88603)} & 8.54908 & {\color[HTML]{CC0000} (+4.74965)} & 45\%↓ & {\color[HTML]{333333} 0.68} & 0.55 & {\color[HTML]{000000} 0.56} & {\color[HTML]{0070C0} (+0.01)} & {\color[HTML]{000000} 0.32} & {\color[HTML]{CC0000} (-0.24)} & 0.25↑ \\
Bayesian Ridge & 3.46493 & {\color[HTML]{333333} 3.69867} & 4.44005 & {\color[HTML]{CC0000} (+0.74139)} & 8.35443 & {\color[HTML]{CC0000} (+4.65577)} & 47\%↓ & {\color[HTML]{333333} 0.64} & 0.58 & {\color[HTML]{000000} 0.61} & {\color[HTML]{0070C0} (+0.03)} & {\color[HTML]{000000} 0.35} & {\color[HTML]{CC0000} (-0.23)} & 0.26↑ \\
SVR & 2.94350 & {\color[HTML]{333333} 3.79201} & 4.42570 & {\color[HTML]{CC0000} (+0.63369)} & 8.25889 & {\color[HTML]{CC0000} (+4.46688)} & 46\%↓ & {\color[HTML]{333333} 0.74} & 0.55 & {\color[HTML]{000000} 0.61} & {\color[HTML]{0070C0} (+0.06)} & {\color[HTML]{000000} 0.36} & {\color[HTML]{CC0000} (-0.19)} & 0.25↑ \\
KNNR & \textbf{1.51625} & {\color[HTML]{333333} 3.73914} & \textbf{4.39229} & {\color[HTML]{CC0000} (+0.65315)} & \textbf{8.09227} & {\color[HTML]{CC0000} (+4.35314)} & 46\%↓ & {\color[HTML]{333333} 0.89} & 0.57 & {\color[HTML]{000000} \textbf{0.62}} & {\color[HTML]{0070C0} (+0.05)} & {\color[HTML]{000000} \textbf{0.39}} & {\color[HTML]{CC0000} (-0.18)} & 0.23↑ \\
Random Forest & 1.87816 & {\color[HTML]{333333} 3.69857} & 4.68172 & {\color[HTML]{CC0000} (+0.98315)} & 8.48477 & {\color[HTML]{CC0000} (+4.78620)} & 45\%↓ & {\color[HTML]{333333} 0.89} & 0.57 & {\color[HTML]{000000} 0.56} & {\color[HTML]{CC0000} (-0.01)} & {\color[HTML]{000000} 0.33} & {\color[HTML]{CC0000} (-0.25)} & 0.24↑ \\
Gradient Boosting & 1.85019 & {\color[HTML]{333333} 3.55563} & 4.59349 & {\color[HTML]{CC0000} (+1.03786)} & 8.50953 & {\color[HTML]{CC0000} (+4.95389)} & 46\%↓ & {\color[HTML]{333333} \textbf{0.90}} & 0.61 & {\color[HTML]{000000} 0.58} & {\color[HTML]{CC0000} (-0.03)} & {\color[HTML]{000000} 0.32} & {\color[HTML]{CC0000} (-0.28)} & 0.26↑ \\
LightGBM & 1.90642 & {\color[HTML]{333333} 3.64394} & 4.65900 & {\color[HTML]{CC0000} (+1.01506)} & 8.54651 & {\color[HTML]{CC0000} (+4.90258)} & 45\%↓ & {\color[HTML]{333333} 0.89} & 0.59 & {\color[HTML]{000000} 0.57} & {\color[HTML]{CC0000} (-0.02)} & {\color[HTML]{000000} 0.32} & {\color[HTML]{CC0000} (-0.27)} & 0.25↑ \\
TPOT & 2.03313 & {\color[HTML]{333333} \textbf{3.52637}} & 4.64980 & {\color[HTML]{CC0000} (+1.12343)} & 8.53731 & {\color[HTML]{CC0000} (+5.01094)} & 46\%↓ & {\color[HTML]{333333} 0.88} & \textbf{0.61} & {\color[HTML]{000000} 0.57} & {\color[HTML]{CC0000} (-0.04)} & {\color[HTML]{000000} 0.32} & {\color[HTML]{CC0000} (-0.29)} & 0.25↑ \\ \hline
\textit{Mean} & 2.35925 & 3.68172 & 4.56594 & {\color[HTML]{CC0000} (+0.88422)} & 8.41660 & {\color[HTML]{CC0000} (+4.73488)} & 46\%↓ & 0.81 & 0.58 & {\color[HTML]{000000} 0.58} & {\color[HTML]{0070C0} (+0.01)} & {\color[HTML]{000000} 0.34} & {\color[HTML]{CC0000} (-0.24)} & 0.25↑ \\ \hline
\end{tabular}
}
    \caption{Glass transition temperature ($T_{g}$)}
    \label{subtable:1}
  \end{subtable}
\end{center}


\begin{center}
  \begin{subtable}{\textwidth}
    \centering
    \small
    \resizebox{0.95\textwidth}{!}{
    \centering
    \begin{tabular}{lcccccccccccccc}
\hline
\multicolumn{1}{c}{} & \multicolumn{7}{c}{RMSE} & \multicolumn{7}{c}{$R^{2}$} \\ \cline{2-15} 
\multicolumn{1}{c}{} & \multicolumn{1}{c}{} & \multicolumn{1}{c}{} & \multicolumn{5}{c}{Appended test set} & \multicolumn{1}{c}{} & \multicolumn{1}{c}{} & \multicolumn{5}{c}{Appended test set} \\ \cline{4-8} \cline{11-15} 
\multicolumn{1}{c}{\multirow{-3}{*}{Model}} & \multicolumn{1}{c}{\multirow{-2}{*}{Training set}} & \multicolumn{1}{c}{\multirow{-2}{*}{Test set}} & \multicolumn{2}{c}{w/ re-expt. ($\mathcal{\overline{D}}'$)} & \multicolumn{2}{c}{w/ outliers ($\mathcal{D}'$)} & \multicolumn{1}{l}{$\mathcal{\overline{D}}$ vs. $\mathcal{D}'$} & \multicolumn{1}{c}{\multirow{-2}{*}{Training set}} & \multicolumn{1}{c}{\multirow{-2}{*}{Test set}} & \multicolumn{2}{c}{w/ re-expt. ($\mathcal{\overline{D}}'$)} & \multicolumn{2}{c}{w/ outliers ($\mathcal{D}'$)} & \multicolumn{1}{l}{$\mathcal{\overline{D}}'$ vs. $\mathcal{D}'$} \\ \hline
Elastic Net & 0.03947 & {\color[HTML]{000000} 0.05141} & 0.05158 & {\color[HTML]{CC0000} (+0.00017)} & 0.06641 & {\color[HTML]{CC0000} (+0.01500)} & 22\%↓ & {\color[HTML]{333333} 0.79} & {\color[HTML]{000000} 0.64} & {\color[HTML]{000000} 0.69} & {\color[HTML]{0070C0} (+0.04)} & {\color[HTML]{000000} 0.63} & {\color[HTML]{CC0000} (-0.01)} & 0.05↑ \\
Bayesian Ridge & 0.03923 & {\color[HTML]{000000} 0.04897} & 0.04955 & {\color[HTML]{CC0000} (+0.00058)} & 0.06591 & {\color[HTML]{CC0000} (+0.01694)} & 25\%↓ & {\color[HTML]{333333} 0.79} & {\color[HTML]{000000} 0.69} & {\color[HTML]{000000} 0.72} & {\color[HTML]{0070C0} (+0.03)} & {\color[HTML]{000000} 0.64} & {\color[HTML]{CC0000} (-0.05)} & 0.08↑ \\
SVR & 0.02907 & {\color[HTML]{000000} 0.04489} & 0.04703 & {\color[HTML]{CC0000} (+0.00214)} & 0.06336 & {\color[HTML]{CC0000} (+0.01847)} & 26\%↓ & {\color[HTML]{333333} 0.88} & {\color[HTML]{000000} 0.73} & {\color[HTML]{000000} 0.75} & {\color[HTML]{0070C0} (+0.01)} & {\color[HTML]{000000} 0.67} & {\color[HTML]{CC0000} (-0.07)} & 0.08↑ \\
KNNR & \textbf{0.01005} & {\color[HTML]{000000} 0.04475} & 0.04485 & {\color[HTML]{CC0000} (+0.00011)} & 0.06588 & {\color[HTML]{CC0000} \textbf{(+0.02113)}} & 32\%↓ & {\color[HTML]{333333} \textbf{0.98}} & {\color[HTML]{000000} 0.73} & {\color[HTML]{000000} 0.77} & {\color[HTML]{0070C0} (+0.03)} & {\color[HTML]{000000} 0.64} & {\color[HTML]{CC0000} (-0.09)} & 0.13↑ \\
Random Forest & 0.01948 & {\color[HTML]{000000} 0.04135} & 0.04267 & {\color[HTML]{CC0000} (+0.00131)} & 0.06113 & {\color[HTML]{CC0000} (+0.01978)} & 30\%↓ & {\color[HTML]{333333} 0.95} & {\color[HTML]{000000} 0.78} & {\color[HTML]{000000} 0.79} & {\color[HTML]{0070C0} (+0.01)} & {\color[HTML]{000000} 0.69} & {\color[HTML]{CC0000} (-0.08)} & 0.10↑ \\
Gradient Boosting & 0.02090 & {\color[HTML]{000000} 0.04013} & 0.04194 & {\color[HTML]{CC0000} (+0.00181)} & 0.06037 & {\color[HTML]{CC0000} (+0.02024)} & 31\%↓ & {\color[HTML]{333333} 0.94} & {\color[HTML]{000000} 0.79} & {\color[HTML]{000000} 0.80} & {\color[HTML]{0070C0} (+0.01)} & {\color[HTML]{000000} 0.70} & {\color[HTML]{CC0000} (-0.09)} & 0.10↑ \\
LightGBM & 0.02187 & {\color[HTML]{000000} 0.04010} & 0.04206 & {\color[HTML]{CC0000} (+0.00196)} & \textbf{0.05983} & {\color[HTML]{CC0000} (+0.01972)} & 30\%↓ & {\color[HTML]{333333} 0.93} & {\color[HTML]{000000} 0.79} & {\color[HTML]{000000} 0.80} & {\color[HTML]{0070C0} (+0.01)} & {\color[HTML]{000000} \textbf{0.70}} & {\color[HTML]{CC0000} (-0.08)} & 0.09↑ \\
TPOT & 0.01949 & {\color[HTML]{000000} \textbf{0.03971}} & \textbf{0.04165} & {\color[HTML]{CC0000} (+0.00194)} & 0.06026 & {\color[HTML]{CC0000} (+0.02055)} & 31\%↓ & {\color[HTML]{333333} 0.95} & {\color[HTML]{000000} \textbf{0.79}} & {\color[HTML]{000000} \textbf{0.80}} & {\color[HTML]{0070C0} (+0.01)} & {\color[HTML]{000000} 0.70} & {\color[HTML]{CC0000} (-0.09)} & 0.10↑ \\ \hline
\textit{Mean} & 0.02495 & {\color[HTML]{000000} 0.04391} & 0.04517 & {\color[HTML]{CC0000} (+0.00125)} & 0.06289 & {\color[HTML]{CC0000} (+0.01898)} & 28\%↓ & 0.90 & {\color[HTML]{000000} 0.74} & {\color[HTML]{000000} 0.76} & {\color[HTML]{0070C0} (+0.02)} & {\color[HTML]{000000} 0.67} & {\color[HTML]{CC0000} (-0.07)} & 0.09↑ \\ \hline
\end{tabular}
}
    \caption{Tan $\delta$ peak}
    \label{subtable:2}
  \end{subtable}
\end{center}


\begin{center}
  \begin{subtable}{\textwidth}
    \centering
    \small
    \resizebox{0.95\textwidth}{!}{
    \centering
    \begin{tabular}{lcccccccccccccc}
\hline
\multicolumn{1}{c}{} & \multicolumn{7}{c}{RMSE} & \multicolumn{7}{c}{$R^{2}$} \\ \cline{2-15} 
\multicolumn{1}{c}{} & \multicolumn{1}{c}{} & \multicolumn{1}{c}{} & \multicolumn{5}{c}{Appended test set} & \multicolumn{1}{c}{} & \multicolumn{1}{c}{} & \multicolumn{5}{c}{Appended test set} \\ \cline{4-8} \cline{11-15} 
\multicolumn{1}{c}{\multirow{-3}{*}{Model}} & \multicolumn{1}{c}{\multirow{-2}{*}{Training set}} & \multicolumn{1}{c}{\multirow{-2}{*}{Test set}} & \multicolumn{2}{c}{w/ re-expt. ($\mathcal{\overline{D}}'$)} & \multicolumn{2}{c}{w/ outliers ($\mathcal{D}'$)} & \multicolumn{1}{l}{$\mathcal{\overline{D}}$ vs. $\mathcal{D}'$} & \multicolumn{1}{c}{\multirow{-2}{*}{Training set}} & \multicolumn{1}{c}{\multirow{-2}{*}{Test set}} & \multicolumn{2}{c}{w/ re-expt. ($\mathcal{\overline{D}}'$)} & \multicolumn{2}{c}{w/ outliers ($\mathcal{D}'$)} & \multicolumn{1}{l}{$\mathcal{\overline{D}}'$ vs. $\mathcal{D}'$} \\ \hline
Elastic Net & 0.00020 & {\color[HTML]{000000} 0.00022} & 0.00025 & {\color[HTML]{CC0000} (+0.00003)} & 0.00060 & {\color[HTML]{CC0000} (+0.00039)} & 59\%↓ & {\color[HTML]{333333} 0.70} & {\color[HTML]{000000} 0.64} & {\color[HTML]{000000} 0.60} & {\color[HTML]{CC0000} (-0.04)} & {\color[HTML]{000000} 0.29} & {\color[HTML]{CC0000} (-0.35)} & 0.31↑ \\
Bayesian Ridge & 0.00020 & {\color[HTML]{000000} 0.00022} & 0.00025 & {\color[HTML]{CC0000} (+0.00003)} & 0.00060 & {\color[HTML]{CC0000} (+0.00038)} & 59\%↓ & {\color[HTML]{333333} 0.70} & {\color[HTML]{000000} 0.64} & {\color[HTML]{000000} 0.59} & {\color[HTML]{CC0000} (-0.04)} & {\color[HTML]{000000} 0.29} & {\color[HTML]{CC0000} (-0.34)} & 0.30↑ \\
SVR & 0.00025 & {\color[HTML]{000000} 0.00030} & 0.00033 & {\color[HTML]{CC0000} (+0.00002)} & 0.00064 & {\color[HTML]{CC0000} (+0.00033)} & 49\%↓ & {\color[HTML]{333333} 0.53} & {\color[HTML]{000000} 0.30} & {\color[HTML]{000000} 0.30} & {\color[HTML]{CC0000} (-0.01)} & {\color[HTML]{000000} 0.21} & {\color[HTML]{CC0000} (-0.10)} & 0.09↑ \\
KNNR & \textbf{0.00004} & {\color[HTML]{000000} 0.00024} & 0.00026 & {\color[HTML]{CC0000} (+0.00001)} & 0.00063 & {\color[HTML]{CC0000} \textbf{(+0.00039)}} & 60\%↓ & {\color[HTML]{333333} \textbf{0.99}} & {\color[HTML]{000000} 0.56} & {\color[HTML]{000000} 0.57} & {\color[HTML]{0070C0} (+0.01)} & {\color[HTML]{000000} 0.23} & {\color[HTML]{CC0000} (-0.33)} & 0.34↑ \\
Random Forest & 0.00014 & {\color[HTML]{000000} 0.00022} & 0.00025 & {\color[HTML]{CC0000} (+0.00003)} & 0.00061 & {\color[HTML]{CC0000} (+0.00039)} & 59\%↓ & {\color[HTML]{333333} 0.86} & {\color[HTML]{000000} 0.64} & {\color[HTML]{000000} 0.60} & {\color[HTML]{CC0000} (-0.04)} & {\color[HTML]{000000} 0.29} & {\color[HTML]{CC0000} (-0.35)} & 0.31↑ \\
Gradient Boosting & 0.00015 & {\color[HTML]{000000} 0.00022} & 0.00025 & {\color[HTML]{CC0000} (+0.00003)} & 0.00060 & {\color[HTML]{CC0000} (+0.00038)} & 59\%↓ & {\color[HTML]{333333} 0.82} & {\color[HTML]{000000} 0.64} & {\color[HTML]{000000} 0.60} & {\color[HTML]{CC0000} (-0.04)} & {\color[HTML]{000000} 0.30} & {\color[HTML]{CC0000} (-0.35)} & 0.30↑ \\
LightGBM & 0.00016 & {\color[HTML]{000000} \textbf{0.00022}} & \textbf{0.00025} & {\color[HTML]{CC0000} (+0.00003)} & \textbf{0.00059} & {\color[HTML]{CC0000} (+0.00038)} & 59\%↓ & {\color[HTML]{333333} 0.80} & {\color[HTML]{000000} \textbf{0.65}} & {\color[HTML]{000000} \textbf{0.61}} & {\color[HTML]{CC0000} (-0.04)} & {\color[HTML]{000000} \textbf{0.32}} & {\color[HTML]{CC0000} (-0.33)} & 0.28↑ \\
TPOT & 0.00014 & {\color[HTML]{000000} 0.00022} & 0.00025 & {\color[HTML]{CC0000} (+0.00003)} & 0.00060 & {\color[HTML]{CC0000} (+0.00039)} & 59\%↓ & {\color[HTML]{333333} 0.85} & {\color[HTML]{000000} 0.64} & {\color[HTML]{000000} 0.59} & {\color[HTML]{CC0000} (-0.05)} & {\color[HTML]{000000} 0.29} & {\color[HTML]{CC0000} (-0.35)} & 0.30↑ \\ \hline
\textit{Mean} & 0.00016 & {\color[HTML]{000000} 0.00023} & 0.00026 & {\color[HTML]{CC0000} (+0.00003)} & 0.00061 & {\color[HTML]{CC0000} (+0.00038)} & 58\%↓ & 0.78 & {\color[HTML]{000000} 0.59} & {\color[HTML]{000000} 0.56} & {\color[HTML]{CC0000} (-0.03)} & {\color[HTML]{000000} 0.28} & {\color[HTML]{CC0000} (-0.31)} & 0.28↑ \\ \hline
\end{tabular}
}
    \caption{Crosslinking density ($v_{c}$)}
    \label{subtable:3}
  \end{subtable}
\end{center}
\end{table}
\end{landscape}

\begin{figure}[H]
    \centering
    \begin{subfigure}[b]{\textwidth}
        \centering
        \includegraphics[width=0.75\textwidth]{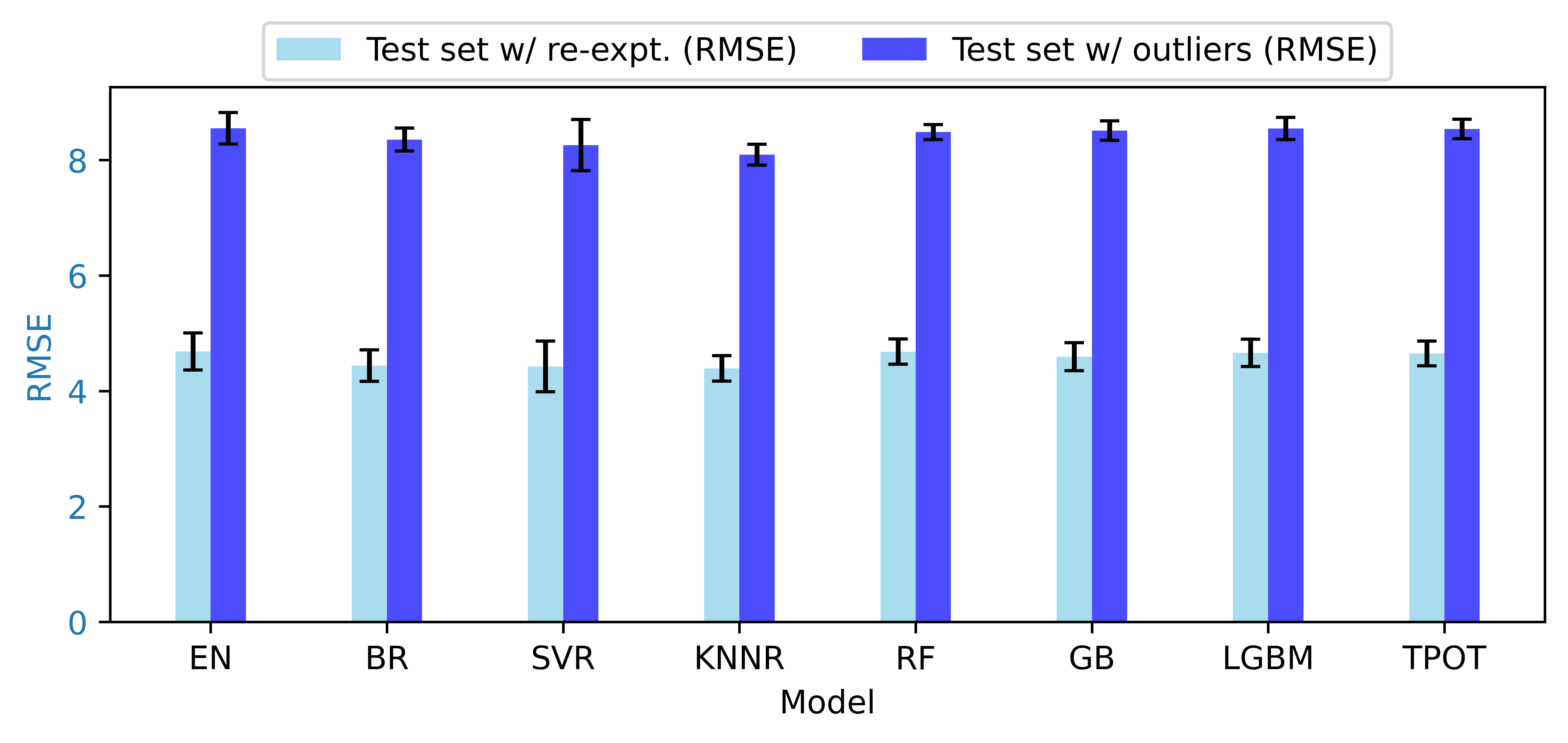}
        \caption{Glass transition temperature ($T_{g}$)}
    \end{subfigure}
    
    \begin{subfigure}[b]{\textwidth}
        \centering
        \includegraphics[width=0.75\textwidth]{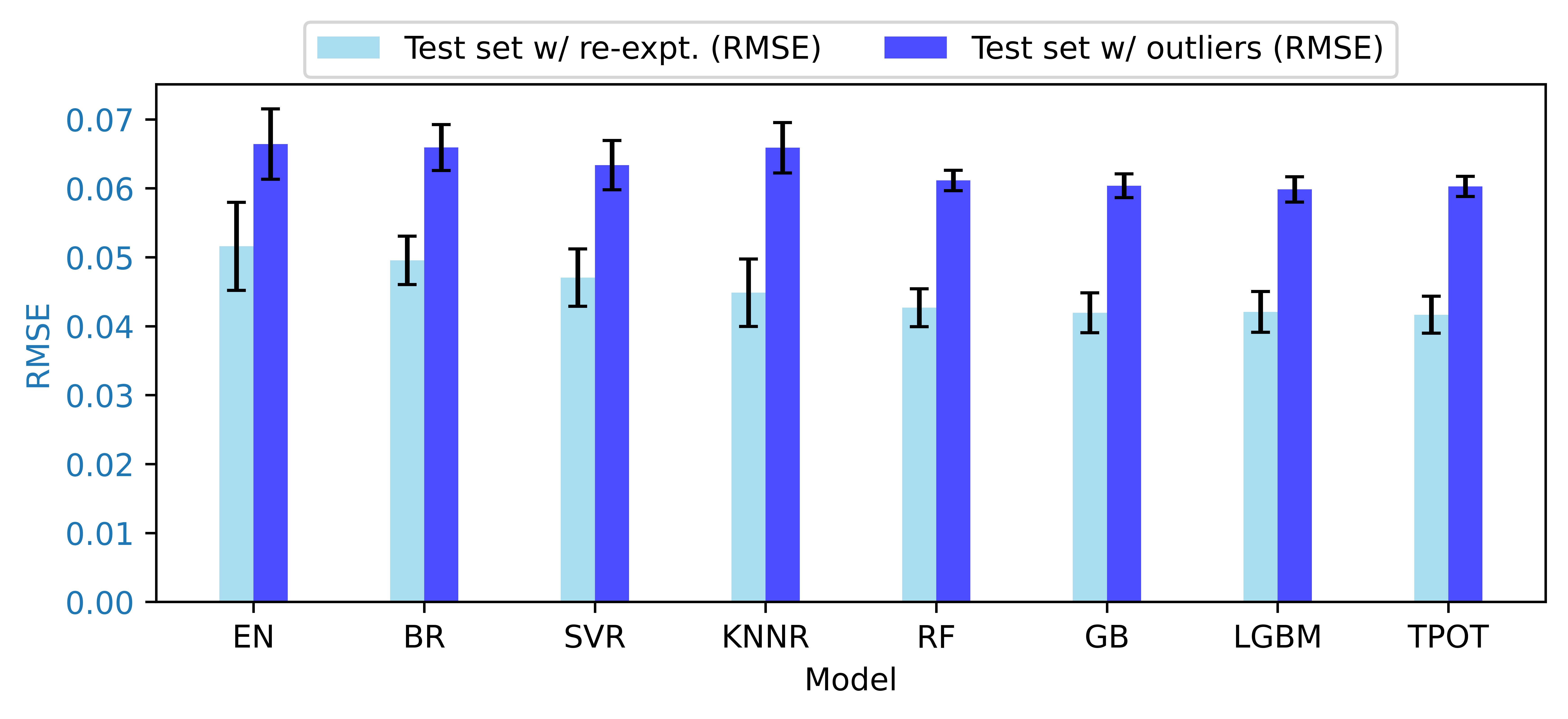}
        \caption{Tan $\delta$ peak}
    \end{subfigure}
    
    \begin{subfigure}[b]{\textwidth}
        \centering
        \includegraphics[width=0.75\textwidth]{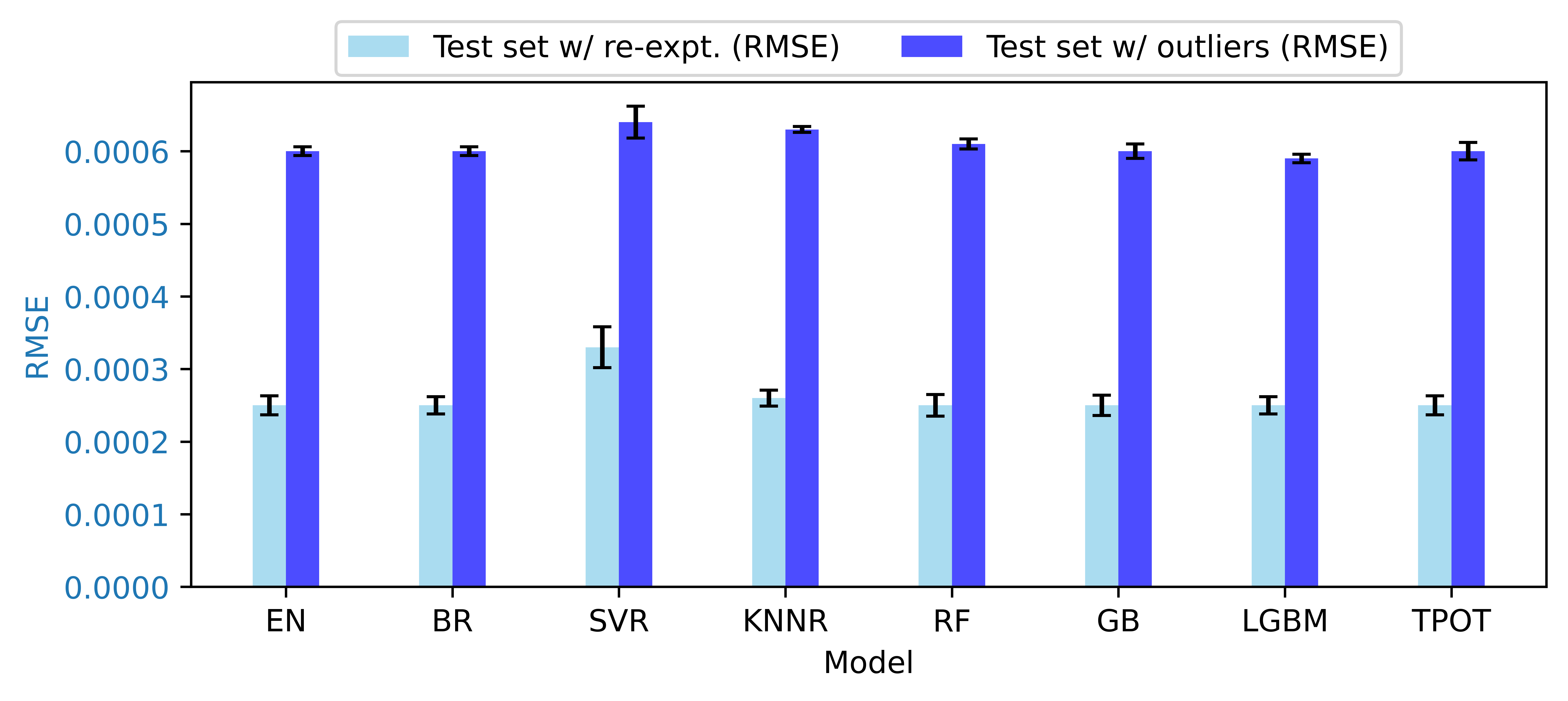}
        \caption{Crosslinking density ($v_{c}$)}
    \end{subfigure}
    \caption{Comparison of root mean squared error (RMSE) across various models for predicting three properties: (a) glass transition temperature ($T_g$), (b) tan $\delta$ peak, and (c) crosslinking density ($v_c$). Each model was trained on the training set and evaluated on two different test sets: one with re-experimental data (light blue bars) and the other with outliers included (dark blue bars).}
    \label{fig:RMSE_histogram}
\end{figure}

\begin{figure}[H]
    \centering
    \begin{subfigure}[b]{\textwidth}
        \centering
        \includegraphics[width=0.75\textwidth]{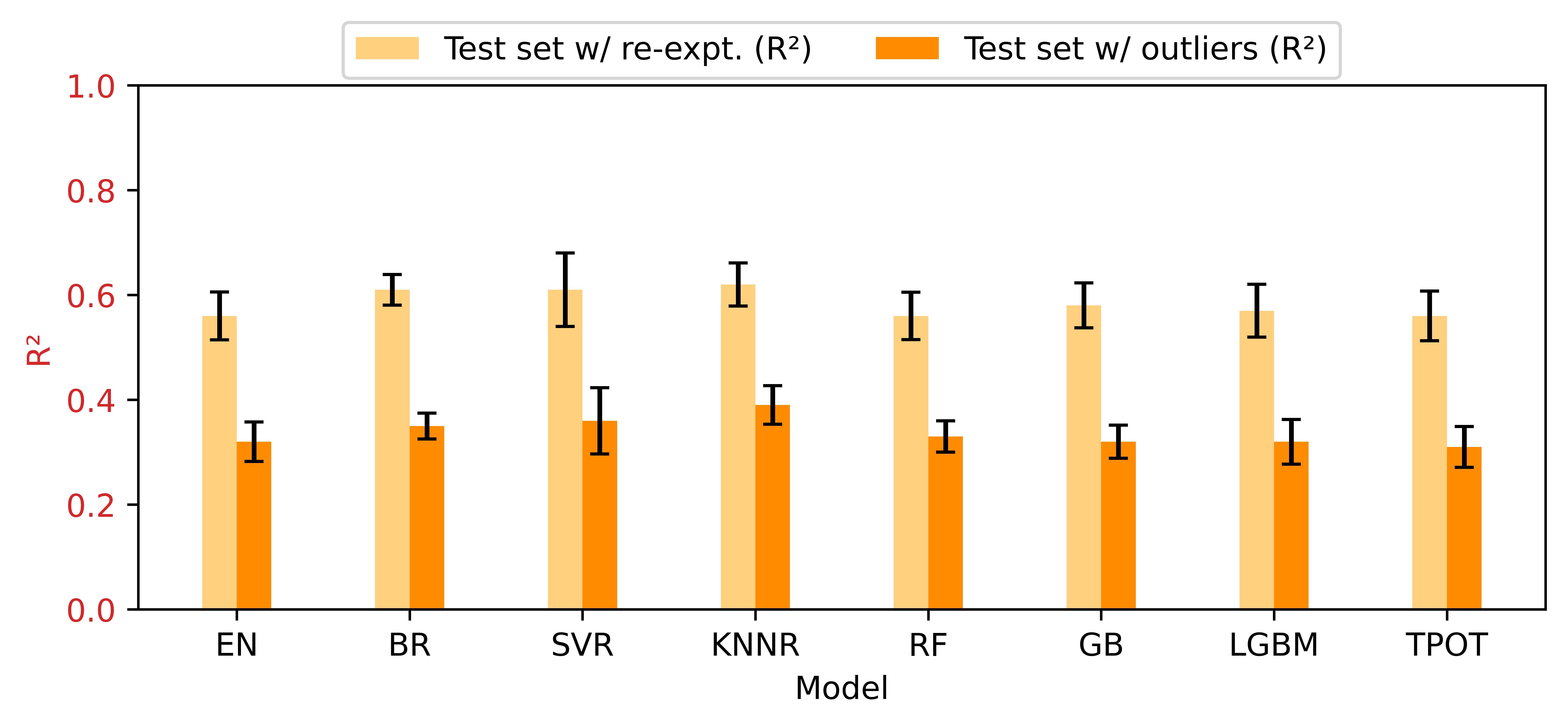}
        \caption{Glass transition temperature ($T_{g}$)}
    \end{subfigure}
    
    \begin{subfigure}[b]{\textwidth}
        \centering
        \includegraphics[width=0.75\textwidth]{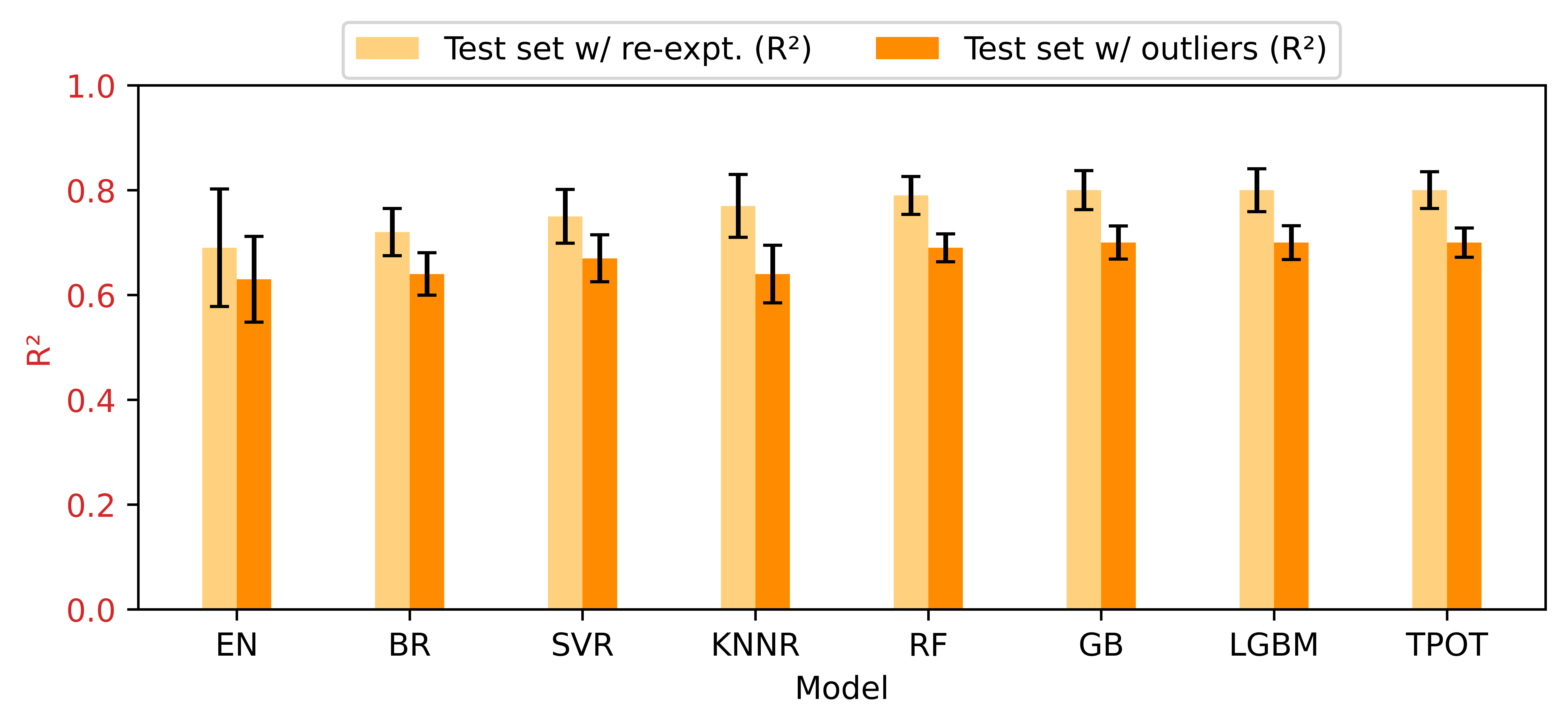}
        \caption{Tan $\delta$ peak}
    \end{subfigure}
    
    \begin{subfigure}[b]{\textwidth}
        \centering
        \includegraphics[width=0.75\textwidth]{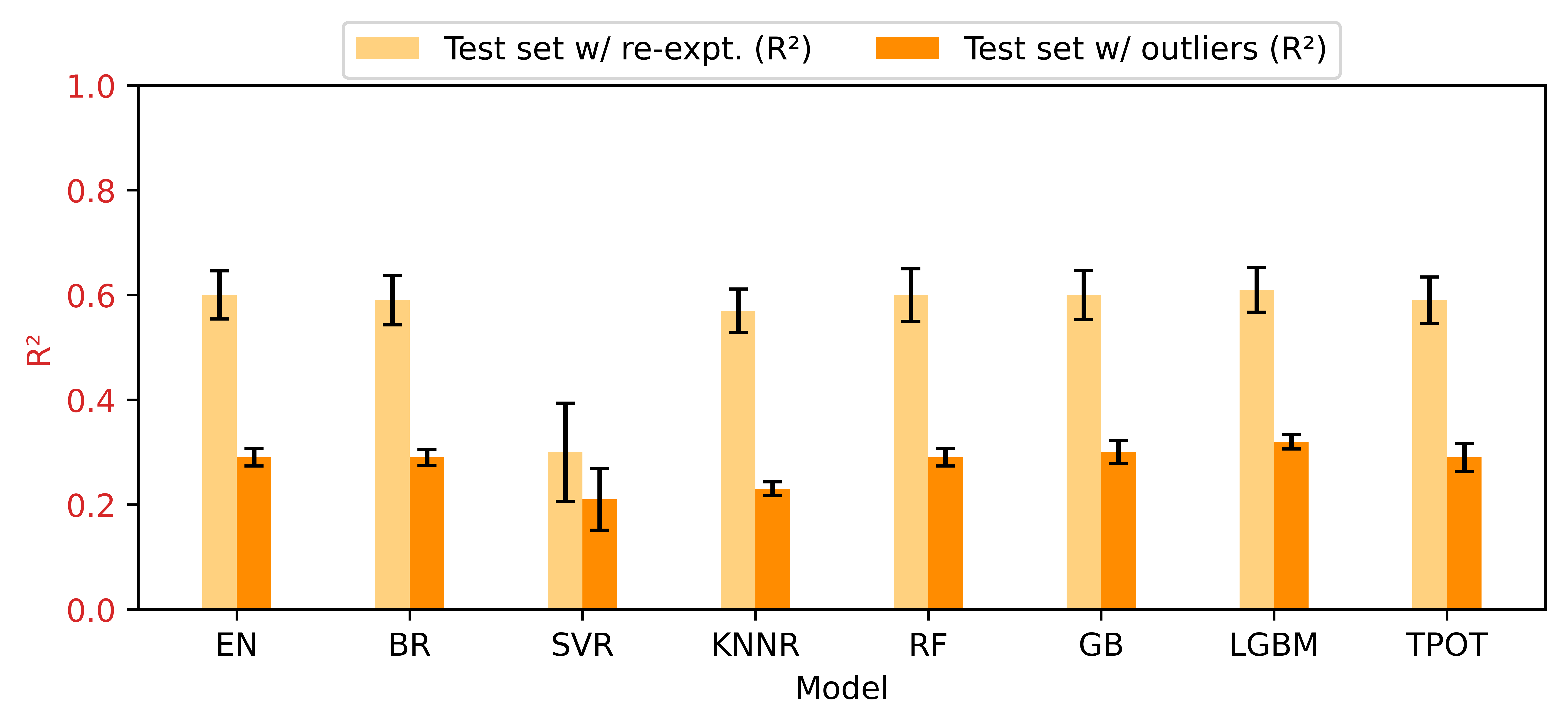}
        \caption{Crosslinking density ($v_{c}$)}
    \end{subfigure}
    \caption{Comparison of coefficient of determination ($R^2$) across various models for predicting three properties: (a) glass transition temperature ($T_g$), (b) tan $\delta$ peak, and (c) crosslinking density ($v_c$). Each model was trained on the training set and evaluated on two different test sets: one with re-experimental data (light orange bars) and the other with outliers included (dark orange bars).}
    \label{fig:R2_histogram}
\end{figure}

\begin{figure}[H]
    \centering
    \begin{subfigure}[b]{0.32\textwidth}
        \centering
        \includegraphics[width=\textwidth]{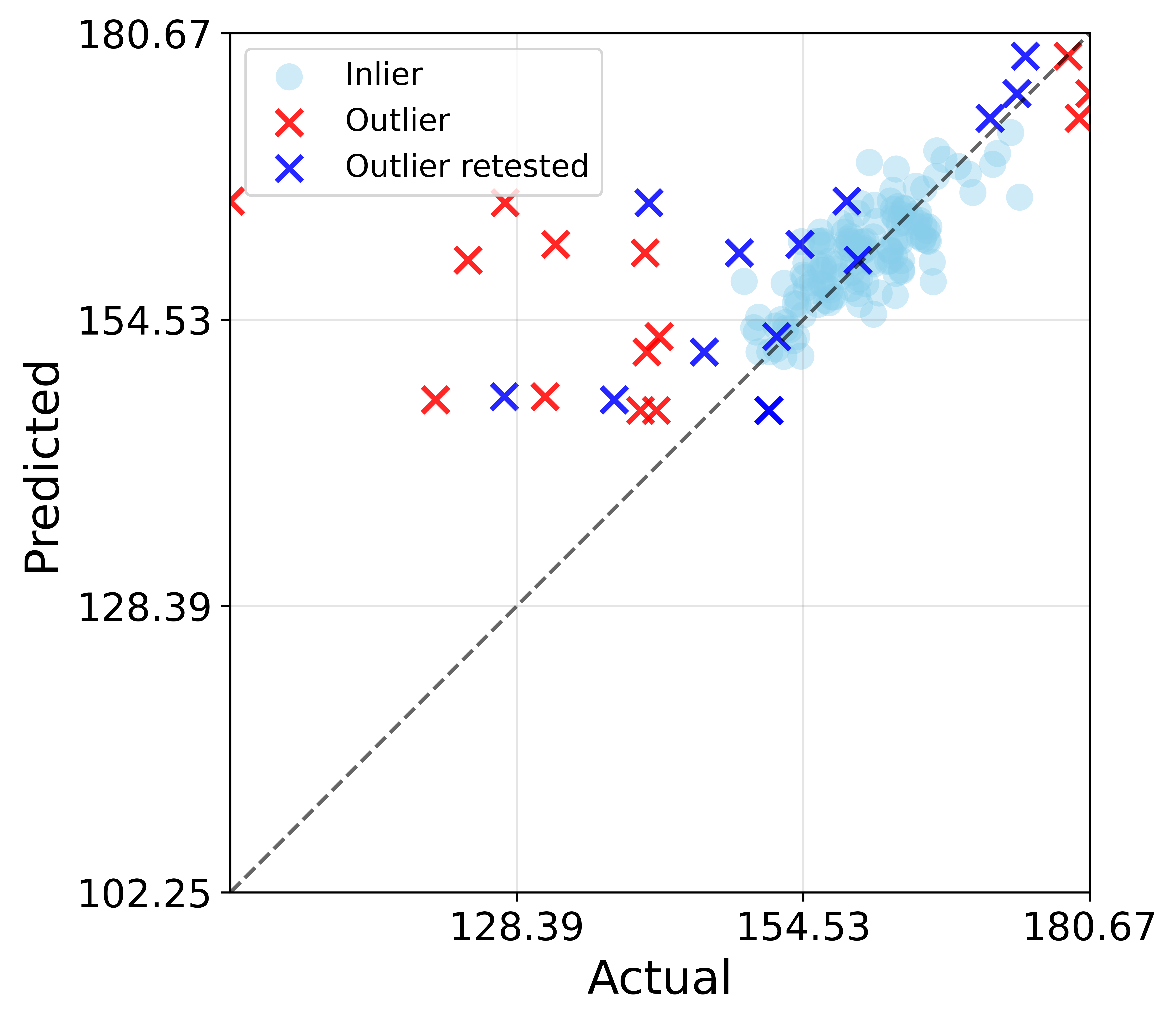}
        \caption{Glass transition temperature ($T_{g}$)}
    \end{subfigure}
    \hfill
    \begin{subfigure}[b]{0.32\textwidth}
        \centering
        \includegraphics[width=\textwidth]{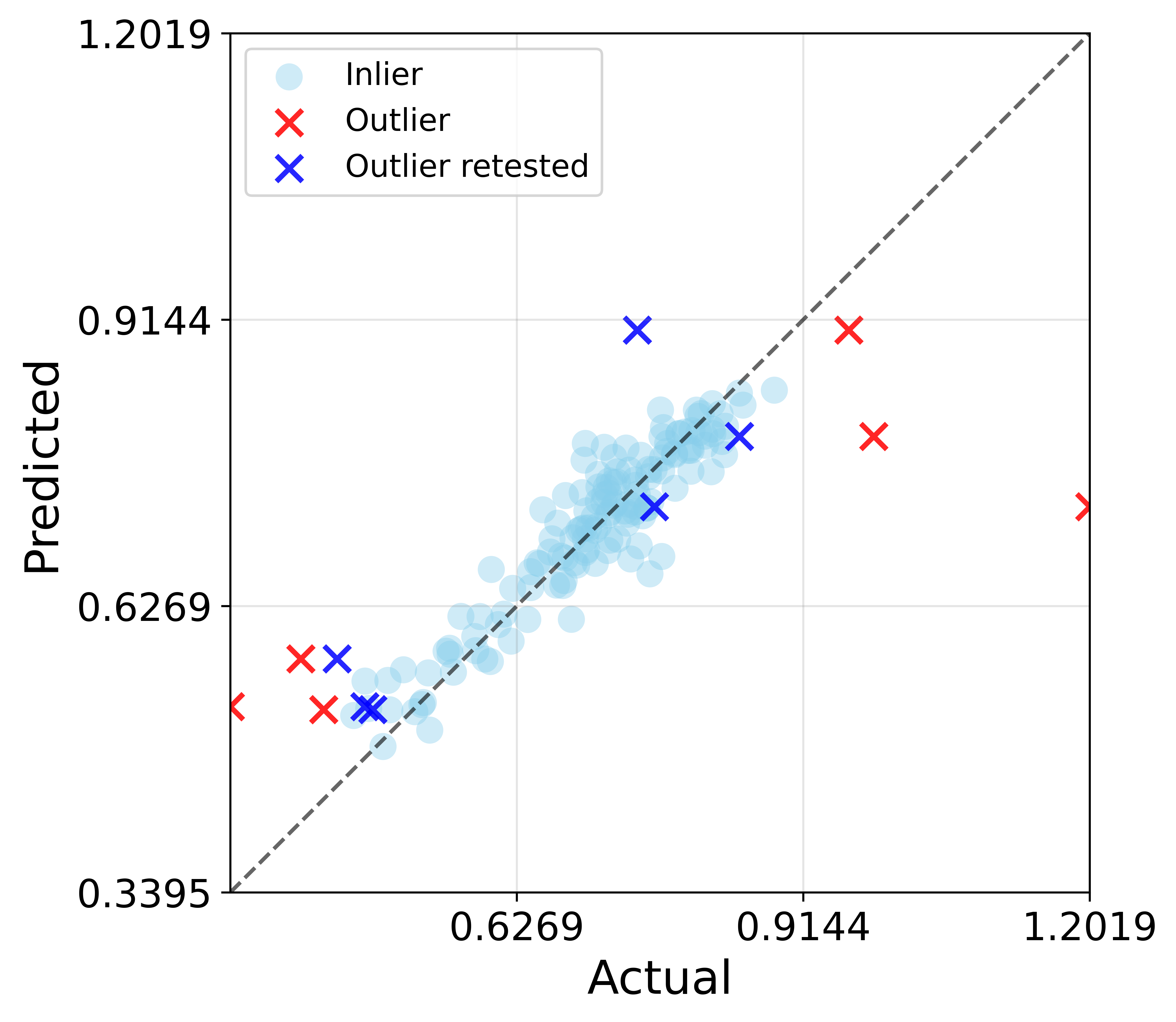}
        \caption{Tan $\delta$ peak}
    \end{subfigure}
    \hfill
    \begin{subfigure}[b]{0.34\textwidth}
        \centering
        \includegraphics[width=\textwidth]{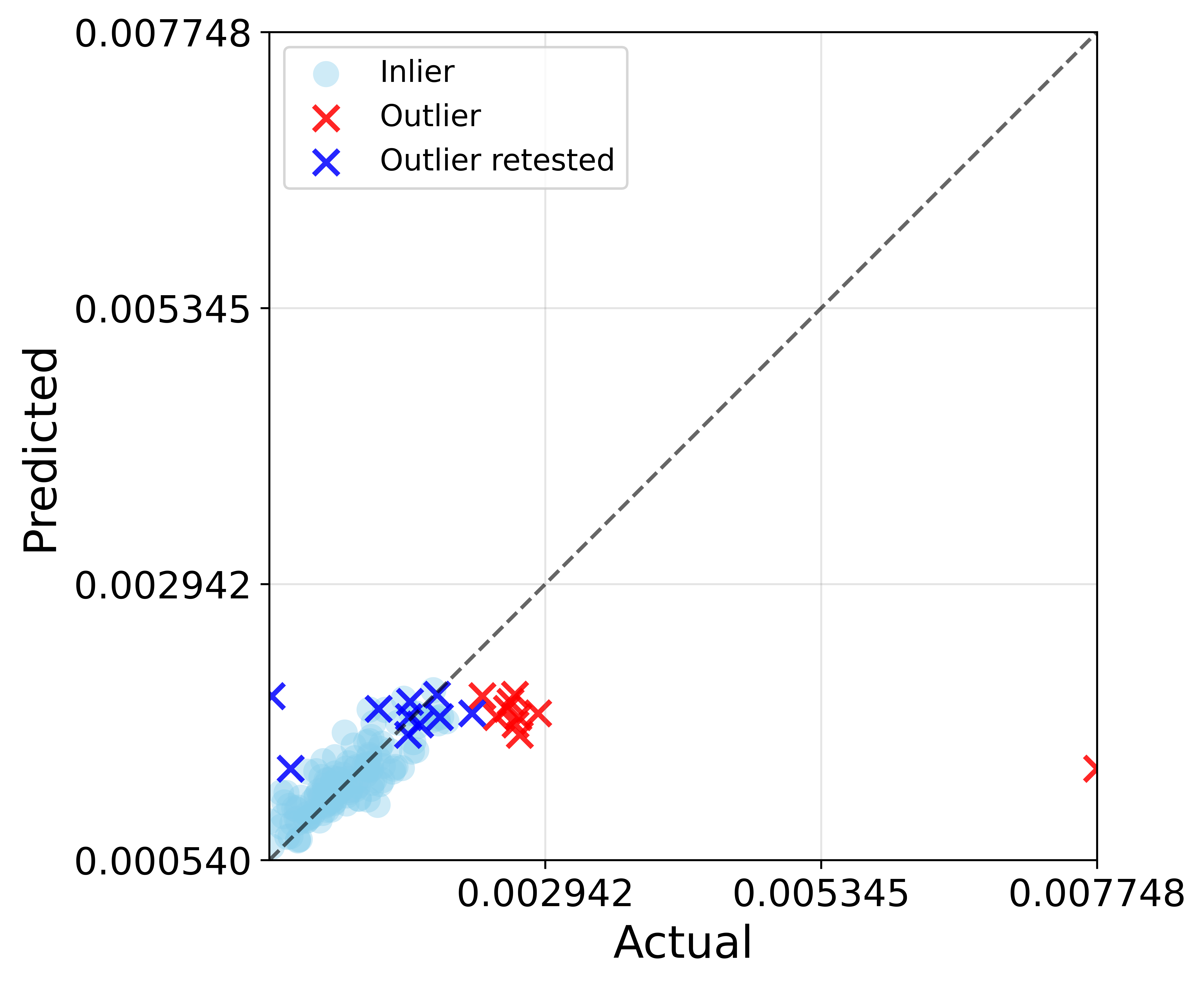}
        \caption{Crosslinking density ($v_{c}$)}
    \end{subfigure}
    \caption{Scatter plots from the TPOT model with the lowest RMSE for each property in the test set, with a random seed of 47 for the train/test split.}
    \label{fig:scatter_plots}
\end{figure}

We further compared datasets with re-experimented data and those containing outliers appended to the existing test set.   
This comparison demonstrates that outliers deviate from the trend of inliers, validating the effectiveness of the outlier detection method. 
As shown in \cref{fig:RMSE_histogram} and \cref{fig:R2_histogram}, including outlier data increases RMSE values and decreases $R^2$ values across all models, while re-experimented data exhibit trends more aligned with the inlier dataset. 
For example, \cref{fig:scatter_plots} highlights these differences using the TPOT model: re-experimented outlier data closely match the inlier data distribution, whereas the original outlier data show significant deviations. 
These results confirm that the outlier detection algorithm effectively identifies data points that significantly diverge from the distribution of the original dataset, suggesting that the detection process was effective.

\subsection{Discussion}
In the proposed re-experimentation approach, the reliability of outlier detection is enhanced by combining three techniques through hard voting to identify outliers for each property from the dataset. 
However, the choice of hyperparameters in outlier detection methods can influence the resulting outliers, necessitating careful mitigation to ensure accurate results. 
A potential direction is to detect outliers in multiple passes of smaller random batches of the dataset, to combine the batchwise results for greater robustness.Alternatively, a multi-step iterative filtering approach, starting with relaxed parameters and gradually tightening them, may reduce false positives and improve detection accuracy. 

Despite its empirical success, this work has limitations due to the dataset's scope, which is confined to a specific family of thermoset epoxy. 
Thus, the proposed method should be validated across a wider range of polymer prediction tasks. 
Additionally, the unsupervised nature of the outlier detection techniques warrants further exploration, as they rely solely on output values, leaving potential improvements from input data unaddressed. 
Thus, multivariate outlier detection approaches that account for interactions and distributions among input variables may offer a more effective method for improving dataset robustness.
Additionally, reverse engineering polymer compositions from their material properties is equally as challenging as directly predicting these properties. 
This could extend into research that employs machine learning models to reverse-engineer potential compound compositions from given property values. 

This study emphasizes the potential to enhance predictive model performance of machine learning in polymer science by improving data quality through selective re-experimentation. 
However, the limitation of data-driven approaches given an extremely small dataset still remains as a significant challenge. 
In scenarios where a limited number of data points can disproportionately influence overall predictions, ensuring data balance and representativeness becomes crucial. 
Future research should consider optimizing sampling strategies during data collection or employing data augmentation techniques to enhance dataset diversity. 
For instance, complementary approaches such as molecular dynamics (MD) simulations or Monte Carlo simulations could be utilized to generate synthetic data, increasing the effective size of the prepared dataset. 
When combined with selective re-experimentation, these strategies have a great potential to build more robust predictive models, ultimately contributing to the reliability and generalizability of machine learning applications in polymer science.

The results of this work highlight the potential of outlier detection to minimize unnecessary re-experimentation, optimizing costs and enhancing the efficiency of epoxy property prediction. 
While our selective re-experimentation improves data quality and model accuracy, manually executing the experiments will continue to introduce unwanted variability.
Hence, a more immediate and practical solution is automating the experiment process itself, ensuring higher consistency across sample preparation, mixing, curing, and property measurement. 
In the future, automating the selection of re-experimentation samples could further refine the automated experiment system, driving another jump in efficiency and precision of empirical polymer science.

\section{Conclusion}

We propose a selective re-experimentation approach to address errors and outliers in experiment-based datasets. 
Our approach makes clever use of outlier detection to choose cases for re-experimentation, and its effectiveness is validated across various machine learning algorithms for predicting the mechanical properties of thermoset epoxy polymers. 
We note a markedly improved dataset quality leads to enhanced predictive performance for the mechanical properties by ameliorating innate experimental variability through selective re-experimentation.
As a result, our method presents a novel cost-effective strategy to bolster dataset quality for downstream machine learning applications in materials science.

Although this study focuses on the mechanical properties of epoxy adhesives, the results open several directions for future research. 
Further validation of this approach on larger datasets from different polymerization systems may demonstrate its broader applicability and reliability. 
Moreover, enhanced datasets could facilitate more complex tasks, such as suggesting polymer compositions to achieve a given set of properties. 
Finally, future research could employ machine learning techniques to reverse-engineer viable compositions, facilitating customized material design and eventually accelerating the development of novel materials.

\bibliography{ref}

\end{document}